\documentclass[12pt]{article}
\usepackage{graphicx}
\usepackage{float}
\usepackage{authblk}

\title{Presence of solar inner F-corona and coronal heating}

\author[*1,2,3]{ Z.Q. Qu}
\author[4]{R.Y. Zhou}
\author[5]{H. Su}
\author[6]{Y. Liang}
\affil[1]{\small Yunnan Observatories, CAS, Tianwentai road, Guandu
district, Kunming, Yunnan, China} \affil[2]{\small School of Space
and Astronomy, Nanjing University, Hankou road, Nanjing, China}
\affil[3]{\small Institute of Optics and Electronics, CAS, Shuangliu
district, Chengdu, China} \affil[4]{\small Department of Physics,
Chinese University of Hong Kong, Shatin, New Territories, Hong Kong
SAR, China} \affil[5]{\small Yunnan Amateur Astronomers Association,
No.2506 Jida Square, Panlong district, Kunming, Yunnan, China}
\affil[6]{\small Shanghai Astronomical Observatory, CAS, Xujiahui
district, Shanghai, China} \affil[*]
{\small Z.Q.Qu: zqqu@ynao.ac.cn}
\date{}

\begin{document}
\maketitle

\begin{abstract}
A new source in solar corona scattering photospheric and
chromospheric Fraunhofer spectral lines is detected below a height
of one solar radius above solar limb, consisting of tenuous and cool
neutral atoms and much fewer once ionized ions. It is demonstrated
via maps at the sample Fraunhofer lines within the band from 516.38
to 539.89nm, reconstructed from one set of spatially successive
raster scanning data. The dataset was obtained from a spectrograph
during the total solar eclipse on April 8, 2024, at Oden, Arkansas,
USA. It is revealed from these maps that both the scattering and its
spatial distribution depend on spectral lines, yielded from
different ionization and excitation states of neutral metal atoms
and ions. The distributions show asymmetry and feature of diffusion
originated from the photosphere and chromosphere. Ratio of the
Fraunhofer line depth to the continuum intensity evaluated over the
observational band peaks at 0.25$\%$ and has an average of 0.32$\%$.
More discrete and weaker diffusion of emission counterparts of some
Fraunhofer lines are detected simultaneously. These properties are
critically different from those owned by that F-corona yielded via
dust grain scattering beyond heights of about two and half solar
radii. Hence a term 'inner F-corona' is dubbed for the assembly of
scattering by this new particle source. It becomes definite now that
the solar corona consists of not only free electrons and ions but
also much fewer yet non-negligible neutral atoms. It is emphasized
that global distributions of the outward neutral atom fluxes and
coronal magnetic loops can make the abnormal Cowling resistance the
most primary mechanism responsible for the coronal heating, via
collisions of the neutral atoms injected with ions in the coronal
loops. This likes the heating process in Tokamak with neutral beam
injection(NBI). Finally, concatenation relevant to the coronal
heating, containing energy supply from magneto-convection, emergence
of magnetic fluxes, the neutral atom fluxes found here as well as
radiative and other losses forms a chain of temperature
self-regulation in the solar atmosphere.
\end{abstract}

{\bf Key Words}\hspace{0.3cm}Sun:corona, Sun: Fraunhofer corona,
Sun: coronal heating

\section{Introduction}

As the outmost atmospheric layer of the sun, solar corona has been
being observed during solar eclipses since ancient times. Giovanni
Cassini coined 'corona' after his observation of 1706 total solar
eclipse. However, it had been regarded to be cooler than its
underneath atmospheric layers, because that it is farther away from
the ultimate solar heat resource-the core. Temperature of the corona
was first reliably diagnosed to be hotter than one million degrees
since Hardness and Young recorded the green coronal line at 530.3nm
during 1869 solar eclipse, and about 65 years later Grotrian
\cite{Grotrian1934} in 1934 and then Edl$\acute{e}$n\cite{Edlen1943}
in 1943 realized that this spectral line is caused by transitions in
thirteen times ionized iron ions(noted as FeXIV) in frame of quantum
theory of atoms. Since then many other forbidden emission lines have
been observed again with high formation temperatures comparable to
that of the green coronal line, the opposite view had been prevailed
that all the coronal particles should be ionized, and the corona
should consist of highly ionized ions and free electrons, with
temperatures up to several million degrees even in the quiet sun
regions\cite{Aschwanden2015}. Therefore, tackling coronal heating
issue has become a great challenge for solar physicists.

Alternatively, attributed to findings of many other lines such as
Extreme Ultra-violet (EUV) ones onboard SOHO\cite{Judge1998} with
their formation temperatures lower than one million degrees, the
above view about the corona is gradually changed. It is regarded now
as a multi-thermal system. For instance, neutral helium atom is
found in the corona(e.g., Stellmacher and
Koutchmy\cite{Stellmacher1974}), judged from its abnormally strong
emission line at 58.43nm, caused by transition 1s$^{2}\rightarrow
1s2p^{1}P^{0}$, as called 'helium enhancement'(e.g., Judge and
Pietarila\cite{Judge2004}). The line formation temperature can be
judged to be about 1.622$\times 10^{5}K$, approximately estimated
from a simple relationship $E=\frac{3}{2}kT$, corresponding to an
excitation energy of 20.964$eV$ for its upper energy level(from {\it
https://www.nist.gov/pml/ atomic-spectra-database}, and hereafter).
In the above expression, $E$ indicates the excitation potential, $T$
the 'formation temperature' and $k(=8.6173\times 10^{-5}eV/K)$ the
Boltzmann coefficient. There are two points of view about the
neutral helium atom origin. One is suggested that they are formed
via recombination of once ionized helium ions with free
electrons\cite{Moise2010}, and the other is that they come from the
photosphere via diffusion across magnetic field canopies
\cite{Judge2004}. However, the latter view has not been supported by
systematic observations of spatial distributions of these neutral
helium atoms.

In solar physics literature, the coronal radiation includes mainly
the forbidden spectral line emission(forming E-corona), attributed
to the highly ionized ions, the continuum irradiance(forming
K-corona), contributed dominantly from free electrons, and the
Fraunhofer(absorption) line depression( forming F-corona). The last
one was recognized to extend from above heights of about two and
half solar radii above solar limb, where dust grains had been found
scattering the Fraunhofer lines originally formed in the
photosphere\cite{Grotrian1934} \cite{Morgan2007}
\cite{Stenborg2021}, till the place where the Zodiacal light is
produced as its outer part. According to space observations, the
dust grain assembly occupies an elliptic space around the
sun\cite{Boe2021} \cite{Burtovoi2022} \cite{Lamy2022}.

Nevertheless, a conscious detection has not yet been accomplished
systematically for presence of other neutral atoms, especially
neutral metal atoms generating spectral lines forming in very low
temperatures, scattering and/or diffracting photospheric and
chromospheric Fraunhofer lines in the inner solar corona, where the
dust grains must be sublimated and repelled to outer space due to
solar irradiance and heat(e.g., Russell\cite{Russell1929}). In fact,
the present work is stimulated from our spectral line polarization
measurement (spectropolarimetry) of very local upper solar
atmosphere during 2013 Gabon total solar eclipse\cite{Qu2024} using
a prototype Fiber Arrayed Solar Optical
Telescope(FASOT)\cite{Qu2011} \cite{Qu2014}. The spectropolarimetry
of recorded Fraunhofer lines reveals local existence of neutral
atoms of metals like iron, magnesium, titanium, calcium, chromium
and scandium as well as some of their once ionized ions in the inner
corona\cite{Qu2024}. The local presence seems to be in form of a
cool cloud in the corona, like that found by Deutsch and
Righini\cite{Deutsch1964} via analyzing the recorded emission CaII K
line obtained during the 1963 total solar eclipse.

\section{Observation}

In order to explore whether there is a global distribution of these
cool neutral metal atoms with low first ionization potentials(FIPs)
smaller than 10$eV$ in the corona, spatial raster scanning of
slit-jaw of a Sol'Ex type spectrograph was performed during 2024
American total solar eclipse on April 8, 2024, on a grassland in
front of a church at Oden town, Arkansas. Its guiding optics has
150mm aperture and slit covers a one-dimensional field of view
greater than four solar radii. The exposure time was set to be 0.1
second and the scanning cadence was 0.2 second. The observational
band spans from 516.38nm to 539.89nm. This band contains neutral
magnesium triplet($b_1$:518.4nm, $b_2$: 517.3nm, and $b_4$:516.7nm),
especially the $b_4$ Fraunhofer line stressed below with an
excitation energy $E$ of 2.71$eV$ of the energy lower level,
corresponding to a formation temperature of 2.98$\times 10^{4}K$.
Two close neutral iron lines respectively at 526.95nm and
527.03nm(hereafter referred to FeI527.0nm) are also contained and
addressed later, with excitation energies respectively of 0.86$eV$
and 1.61$eV$ for their lower levels, corresponding to formation
temperatures of 6.65$\times 10^{3}K$ and 1.24$\times 10^{4}K$.
Otherwise, two close once ionized iron lines at 531.62nm and
531.66nm(hereafter referred to FeII531.7nm) are also emphasized
within the band, with higher formation temperatures of respectively
1.417$\times 10^{5}K$ and 8.55$\times 10^{4}K$, calculated from
ionization and excitation energies 18.32$eV$ and 11.05$eV$
respectively for the lower levels. For comparison, the effective
ionization and excitation potentials of the famous green coronal
line FeXIV530.3nm, also contained in the band, reads as 394.5$eV$,
corresponding to an effective formation temperature of 3.05$\times
10^{6}K$ or $10^{6.48}K$.

During the observation, one entire slit scanning containing 731
spectral frames was successfully performed corresponding to same
number of successive slit positions beyond and across the occulted
solar disk. The scanning took about 2.8 minutes in such a way that
by utilizing the apparent diurnal motion of the sun, tracking via
the equatorial mount was stopped after the slit-jaw being placed
above solar west limb. The dark current and flat fielding were
recorded and reduced in data reduction. Maps from image
reconstruction according to the alignment of the successive slit
positions are obtained for the following analysis.

\section{Data analysis}

Three spectral images are plotted in top panels of Figure 1 after
correction of spectral line bending due to imperfect optics of the
spectrograph. In these spectral images on the top and those spectral
line profiles drawn from them as shown below, dispersion or
wavelength dimension is along the abscissa in each panel, with
wavelength enhancement from left to right(no labels in the top
images but referring to the lower panels). The slit alignment is
along the ordinate. On the top left is the photospheric and
chromospheric Fraunhofer spectral images acquired in the early
partial eclipse phase with the slit placed across the bright solar
disk. Two sample spectral images are plotted of respectively one
slit position above the limb (top middle panel) and the other cross
the blocked solar disk (top right panel) acquired during the
totality. In the latter two spectral images, the green coronal line
at 530.3nm is overwhelmingly bright and the most extensive while the
concerned Fraunhofer lines are much weaker to be seen. In fact, the
ratio of the green coronal line intensity above the adjacent
background continuum to the maximum line depth of the Fraunhofer
lines below the adjacent continuum can reach respectively 63.92 and
68.58 for the sample 40th and 628th slits. The bottom three panels
present the corresponding spectral profiles drawn after integration
over these slit-covered regimes producing the Fraunhofer lines. The
intensities indicated along the ordinates are generally normalized
by their adjacent continuum ones, or they are measured in units of
readout of the detector throughout the paper.

The apparent difference can be easily seen between the reference
spectrum image with only the Fraunhofer lines and the totality
spectral images with prominent emission lines but much weaker
Fraunhofer lines. Let us focus on the variation of relative
intensities among these Fraunhofer lines. This can be examined from
the normalized spectral profiles depicted in the lower panels, in
which emission profiles of the green coronal line are artificially
removed for sake of highlighting the Fraunhofer lines. Although an
overall similarity among the Fraunhofer lines in the lower three
profile panels can be seen, the variation in their relative
intensity exists, like the cases described by our previous
paper\cite{Qu2024}. For instance, our focusing FeII531.7nm lines are
prominent in the reference spectra, while they become evidently
weaker in the spectrum of the slit 628 and even difficult to be
recognized in the spectrum of the slit 40. Similar case can be seen
for other lines like that at 538.3nm. The relative intensity
difference between these groups of lines around 527.0nm and those
around 528.2nm is much greater in the reference spectrum than in the
totality ones. Variation in relative intensities happens also among
these three strong spectral lines around 525nm, as well as among the
six spectral lines around 536.6nm from the reference spectrum to the
totality spectra. Furthermore, such a relative intensity variation
takes place again between these totality spectra. More evidently,
the intensity of the $b_4$ line is comparable to the $b_1$ and $b_2$
ones in the spectrum of the slit 40, but it becomes strikingly
weaker in the spectrum of the slit 628. The two lines around 537.0nm
can be clearly seen in the spectrum of the slit 40 but it becomes
much fainter and relative intensities among them are changed in the
spectrum of the slit 628.

All of these variations described above indicate that the Fraunhofer
line scattering in the corona depends on spectral lines, or resulted
from individual atoms and ions with different ionization and
excitation states rather than solid state of dust grains(cf. \cite{
Russell1929}). It is noteworthy that the Fraunhofer line depth or
depression within the whole observational band is contributed
dominantly from the neutral metal atoms, while the once ionized ions
contribute much less, since much fewer Fraunhofer lines are produced
by the ions within the observational band, and their spectra do not
rank the strongest Fraunhofer lines. The line depths of these
Fraunhofer lines normalized by their adjacent continuum intensities
reach their maximum of only a few percent just above the limb, much
weaker than their original ones formed in the photosphere and
chromosphere, as seen in the reference spectrum. Furthermore, it is
valuable to keep in mind that an appearance of line depression often
means that the Fraunhofer line scattering overwhelms the
corresponding emission along the line-of-sight, and vice versa for
an appearance of line emission.

\begin{figure}
\vspace{-2.6cm}\flushleft \hspace{-1.6cm}
\includegraphics[width=5.1cm,height=8.3cm]{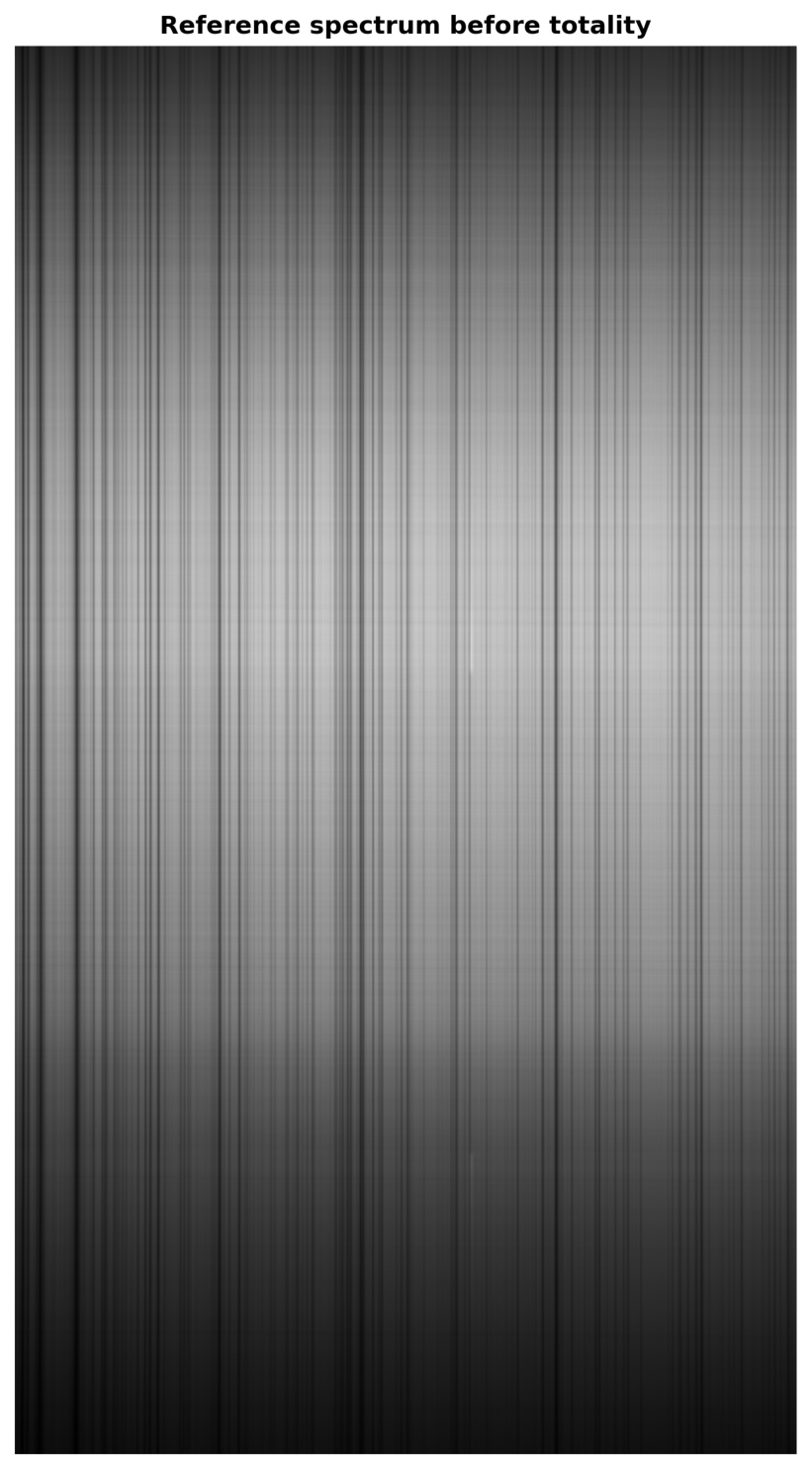}
\hspace{-0.3cm}
\includegraphics[width=5.1cm,height=8.3cm]{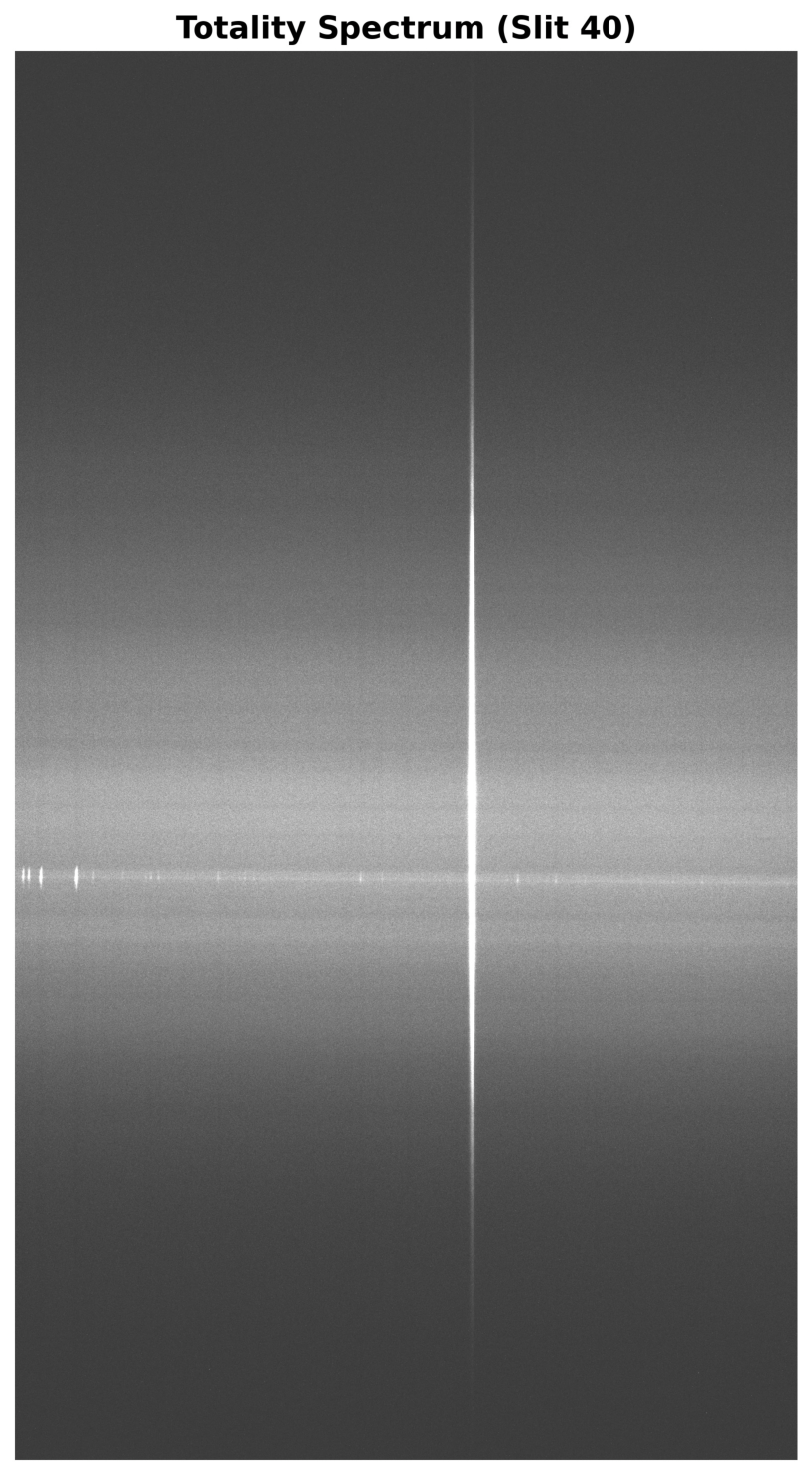}
\hspace{-0.3cm}
\includegraphics[width=5.1cm,height=8.3cm]{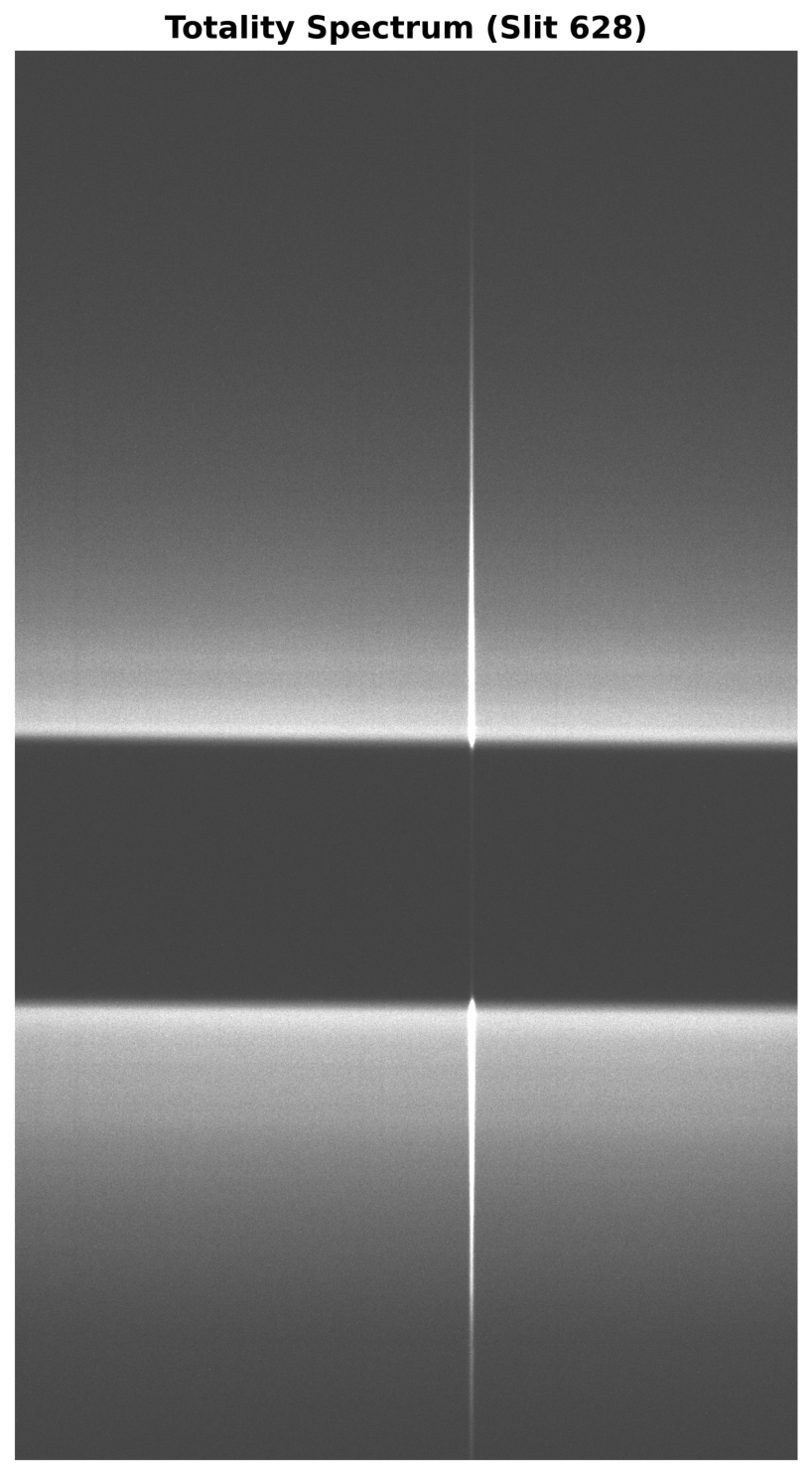}\\
\vspace{-0.1cm}\flushleft\hspace{-1.64cm}
\includegraphics[width=5.06cm,height=3.9cm]{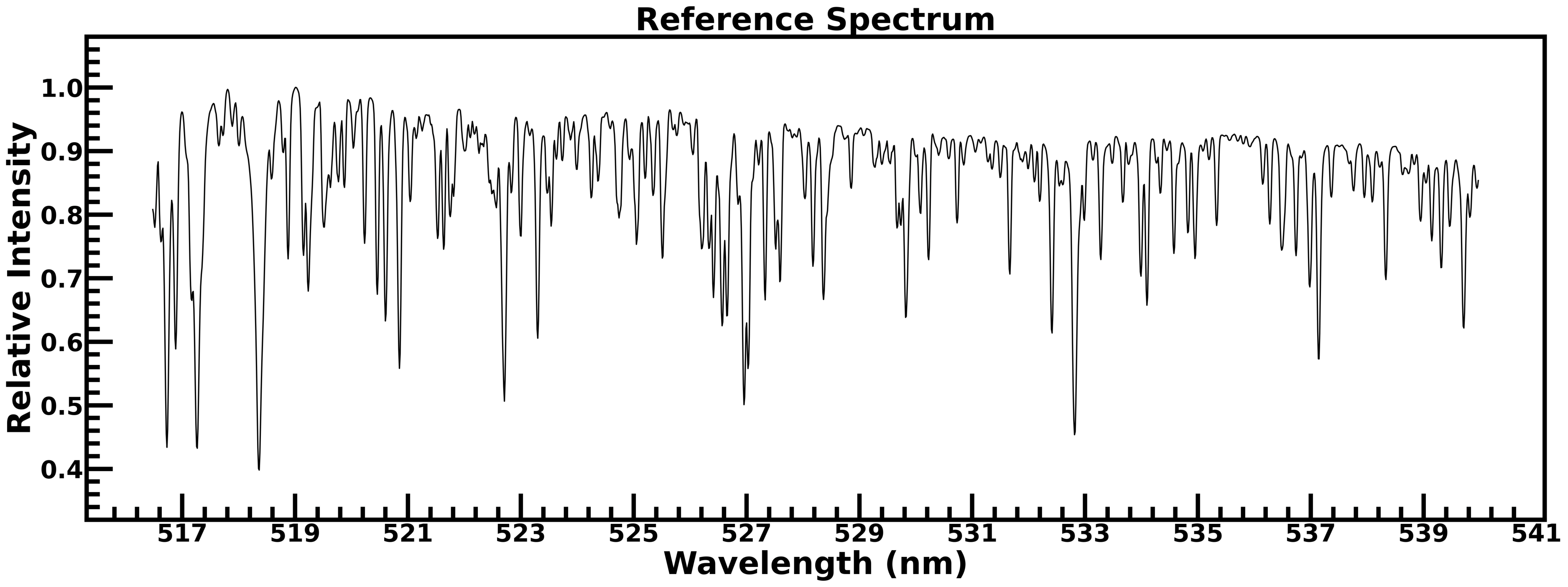}\\
\hspace{-1.83cm}
\includegraphics[width=5.06cm,height=3.9cm]{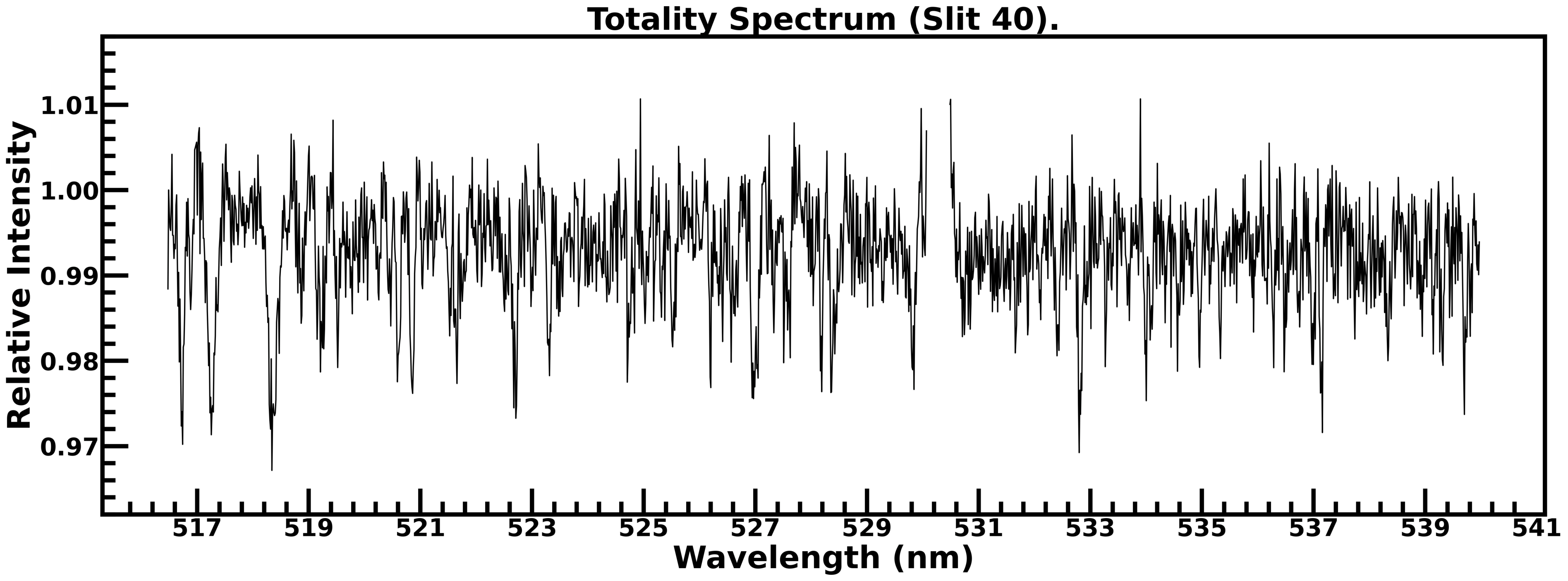}\\
\vspace{0.1cm}\hspace{-1.83cm}
\includegraphics[width=5.06cm,height=3.9cm]{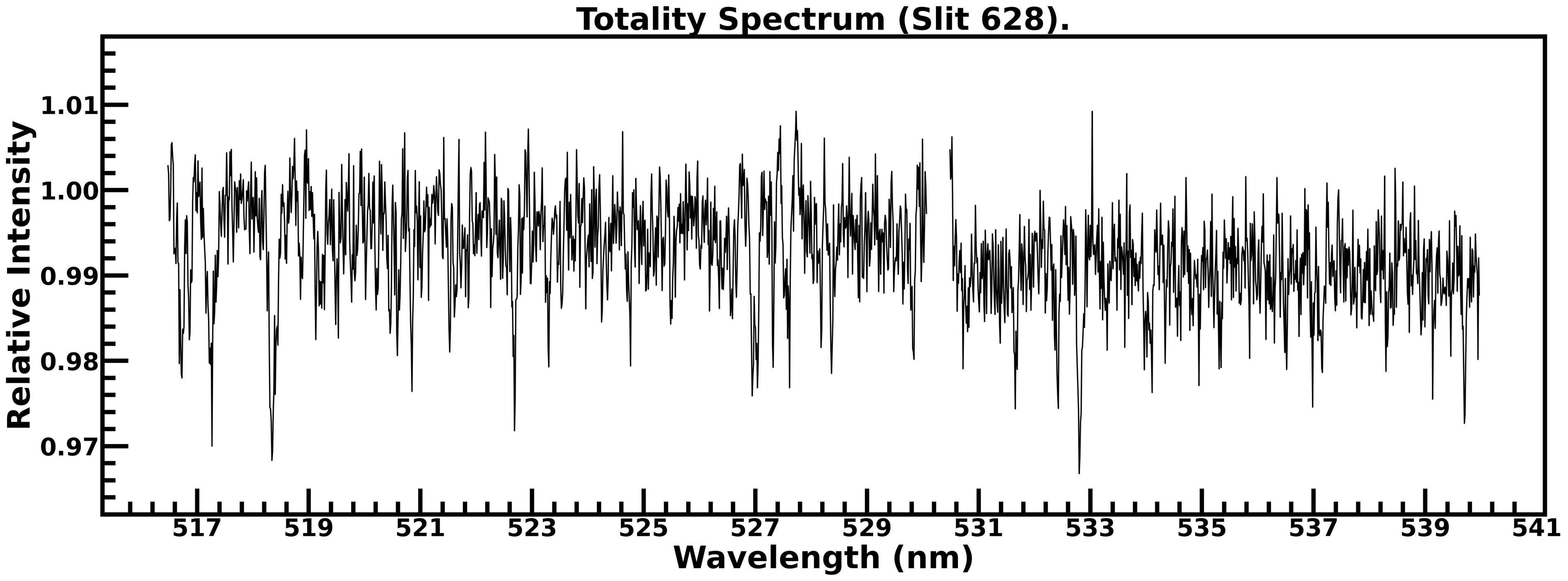}
\caption{\footnotesize Sample spectral images(top panels) and
corresponding profiles(lower panels) drawn from them. The
observational band spans from 516.38nm to 539.89nm, depicted along
the abscissa, and the slit direction is arranged along the ordinate.
The top left one was acquired with the slit crossing bright solar
disk before the total eclipse as the reference spectrum. Spectral
images in the top middle and top right were obtained at the 40th and
628th scanning slit positions respectively above the solar limb,
with lunar disk blocking during the totality. The overwhelmingly
bright vertical spectral line in the two images is the famous green
coronal line(FeXIV530.3nm). In the lower three panels, three
spectral profiles are plotted, drawn from the above spectral images
respectively after the green coronal line profile is removed for
highlighting the Fraunhofer lines. } \label{}
\end{figure}

Information on global scattering distribution of these spatial
points yielding Fraunhofer lines and their counterpart emission
lines can be supplied from maps reconstructed  according to the
spatially successive slit-jaw scanning. We select three
representative groups of the above mentioned Fraunhofer lines and
their emission counterparts as examples. Their relevant maps are
arranged respectively from left to right columns in Figure 2. In
detail, maps of the neutral magnesium $b_4$ Fraunhofer line and its
adjacent continuum are given in the left column panels. Maps of the
two neutral iron atom lines at 527.0nm are plotted in the middle
column. And maps of the two ion lines at 531.7nm are depicted in the
right column. According to their formation temperatures and solar
atmospheric temperature stratification, line cores of these three
groups of Fraunhofer lines are originally formed respectively in the
low chromosphere, the photosphere and the middle chromosphere.
Alternatively, the top row maps of Figure 2 show distributions of
spectral line intensities after integration over all the sample
wavelengths within the line intervals respectively. And maps
reconstructed from their corresponding adjacent continuum
intensities are depicted in panels of the second row(counted from
top to bottom, hereafter). The wavelength range indicated in the
brackets are used to get the average of the continuum intensity.

It is very hard to distinguish the morphologies shown in one map
from another among the maps of the top two rows by direct
inspection. Therefore, to intuitively highlight the F-corona and
distribution of line emission caused from allowed transitions, it is
natural and reasonable to draw maps of difference between the two
top rows respectively for each sample group of lines. In detail, the
integrated line intensity difference used below is quantitatively
obtained from subtraction of the spectral line intensity from their
average adjacent continuum intensity for each sample wavelength
point, and then the differences are integrated over the line
wavelength interval respectively. The difference can be either
negative values reflecting the Fraunhofer line depression or
positive ones representing the line emission. Maps in the bottom two
panels give essentially the Fraunhofer line scattering distribution
as well as their emission counterpart ones. In order to clearly see
the F-corona, those positive values are artificially set to be zero
in the third row maps and to emphasize the counterpart emission
distribution shown in the bottom panels, those negative values are
set to be zero.

Careful view of the third row maps provides us their common features
and differences, though the distribution profiles of the dark and
grey parts look like these above these rows. One of their common
features can be found that the Fraunhofer lines are globally
detectable below a height of one solar radius above the limb.
Another common feature shared by these distributions is overall
clear divergence of the line depression with height. This signifies
the diffusion motion of the scatterers in the corona. Similar
diffusion of neutral helium atoms was also found in inner
heliosphere\cite{Moise2010}. Furthermore, no observable fine
structure in the distribution with fibrils can be found, unlike the
other two kinds of distributions depicted in the top two row panels.
The asymmetric distribution of the line depression is also shared by
all these three distributions, but the detailed difference in the
asymmetry can be witnessed. For instance, the asymmetric spread of
the line depression can be found in these regions around the
north-south poles(the directions are indicated in the middle map of
the second row). The strongest asymmetry occurs for FeI527.0nm
line(see the maps of the third row panel) that the line depression
is much stronger in these southern regions than in those northern
regions. And east-west asymmetry exists also in these three
distributions. The distribution in the west is thicker than in the
east. The most evident asymmetry occurs again in the FeI527.0nm map,
then followed by the FeII531.7nm, and that of the MgI$b_4$ line
shows the weakest east-west asymmetry. On the other hand, these
'monochromatic' F-corona maps delineated by the MgI$b_4$ and
FeII531.7nm lines show that their line depressions are more
concentrated in these regions around the equator(E-W) than in those
around the poles, while the line depression described in the neutral
iron lines at 527.0nm is more concentrated above the lower
hemispheric limb in the map. Otherwise, the line depressions are on
the whole stronger at the neutral atom lines of the MgI$b_4$ and
FeI527.0nm than that at the once ionized iron lines at 531.7nm, as
indicated by the corresponding grey scale bars in the maps. All of
these specify that the global distribution of the line depression
caused by scattering depends on the specific spectral lines, thus
the ionization and excitation states of specific particles. Last but
not the least, the tenuous distribution of the once ionized iron
ions implies that there are magnetic gaps between the
magneto-dominated regions or very weak magnetized domains of plasma
underneath the corona for them to escape from the chromosphere and
transition zone.

All these observational facts stated above from the spectral
analysis and reconstructed maps lead to a conclusion that the
F-corona outlined here is critically different from that F-corona
formed by the dust grains scattering the photospheric and
chromospheric radiation above heights of about two and half solar
radii. The latter can be explained via classical electrodynamical
theory, such as Mie or Rayleigh scattering. Furthermore, no
observation is found that this 'outer' F-corona owns any signature
of the diffusion motion above the solar limb. In fact,
upward(generally observed as blue-shift) motions are popular and
easily found from the Dopplergrams of the photosphere and
chromosphere(e.g., Kumar et al.\cite{Kumar2023}, Yu et al.
\cite{Yu2024}), and even in the transition region and the corona
\cite{Tian2021}. As mentioned above, according to space
observations\cite{Lamy2022} \cite{Burtovoi2022}, this 'outer'
F-corona has a definite shape of an ellipse spread around the sun
symmetrically about the elliptic axes, and extended to the earth's
orbit identified as the Zodiacal light, where the F-corona described
here cannot be detected at the present due to the diffusion.
Therefore, the F-corona detected here is dubbed 'inner F-corona'.

On the other hand, the counterpart line emissions of some Fraunhofer
lines are also detected simultaneously in different spatial points.
The spreads of emission counterparts of these Fraunhofer lines are
also globally observable and diffusive, witnessed in these bottom
maps, but weaker than their line depression counterparts as referred
to the grey scale bars. For instance, line emission in map of the
neutral magnesium $b_4$ line is much weaker than line depression,
especially above the eastern limb. Most of emission points
distributed in scattering appearance can be regarded as noise. On
the contrary, the line emission becomes enhanced well above the
noise close to the eastern limb for the neutral iron lines at
527.0nm, which makes the asymmetry a little different from its
corresponding line depression distribution. The two kinds of maps
look like each other most similarly at the once ionized iron ion
lines at 531.7nm, and even the amplitudes of their line emission and
depression are the closest among these three distributions in the
bottom panels. Evidently, these line emissions are also more
concentrated above the active regions again. Therefore, the presence
of the line emissions especially shown in the bottom middle panel
reinforces the idea that the corona consists of not only the ions
and free electrons but also neutral atoms as its thinnest part. In
other words, neutral corona exists. It is noteworthy that the
diffusion scenarios of both scatterer and emitters provide us a
collateral evidence for diffusion of the neutral helium atoms and
once ionized helium ions into the corona, as suggested by Judge and
Pietarila\cite{Judge2004}.

\begin{figure}
\flushleft\vspace{-3.4cm}\hspace{-1.68cm}
\includegraphics[width=5.0cm,height=6.26cm]{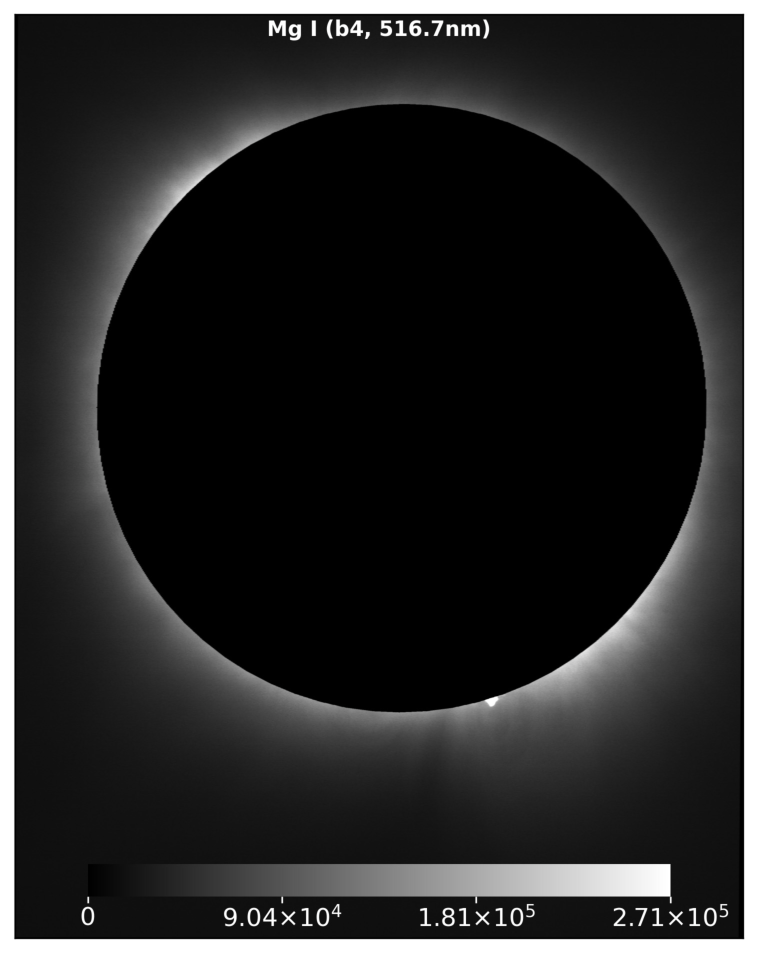}
\hspace{-0.2cm}
\includegraphics[width=5.0cm,height=6.26cm]{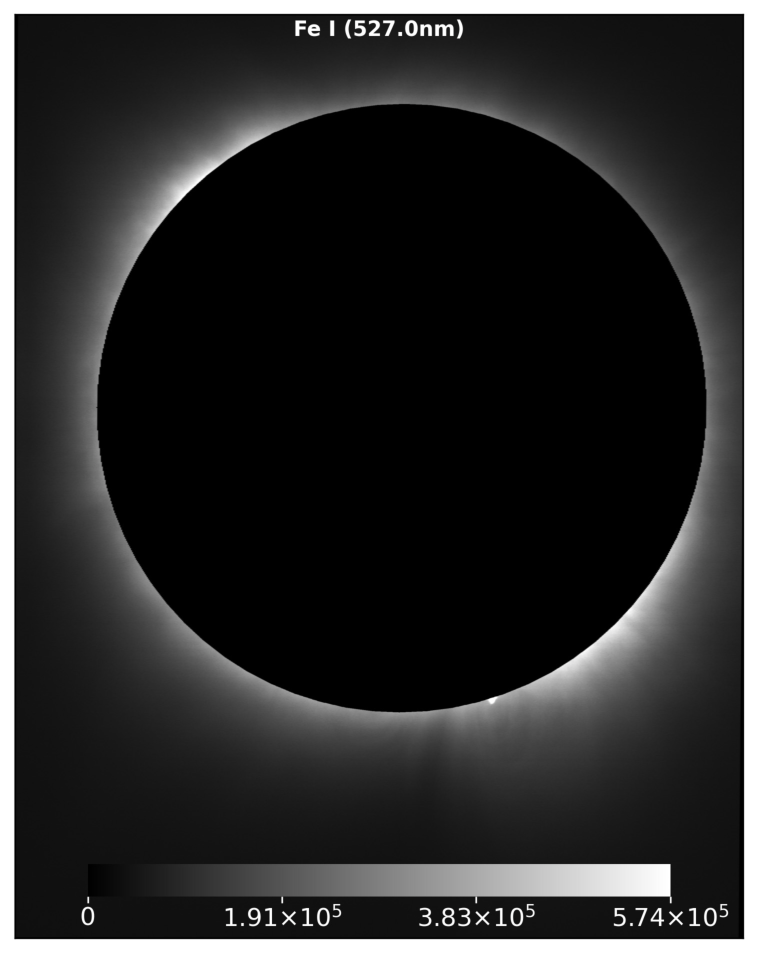}
\hspace{-0.2cm}
\includegraphics[width=5.0cm,height=6.26cm]{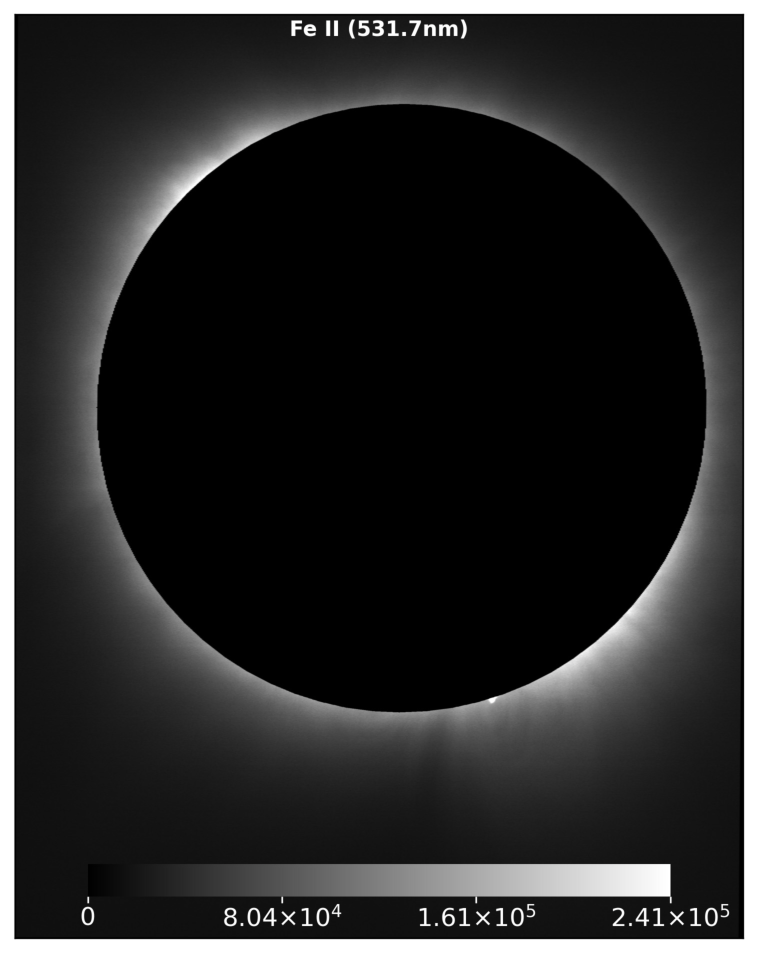}\\
\flushleft\vspace{-0.6cm}\hspace{-1.68cm}
\includegraphics[width=5.0cm,height=6.26cm]{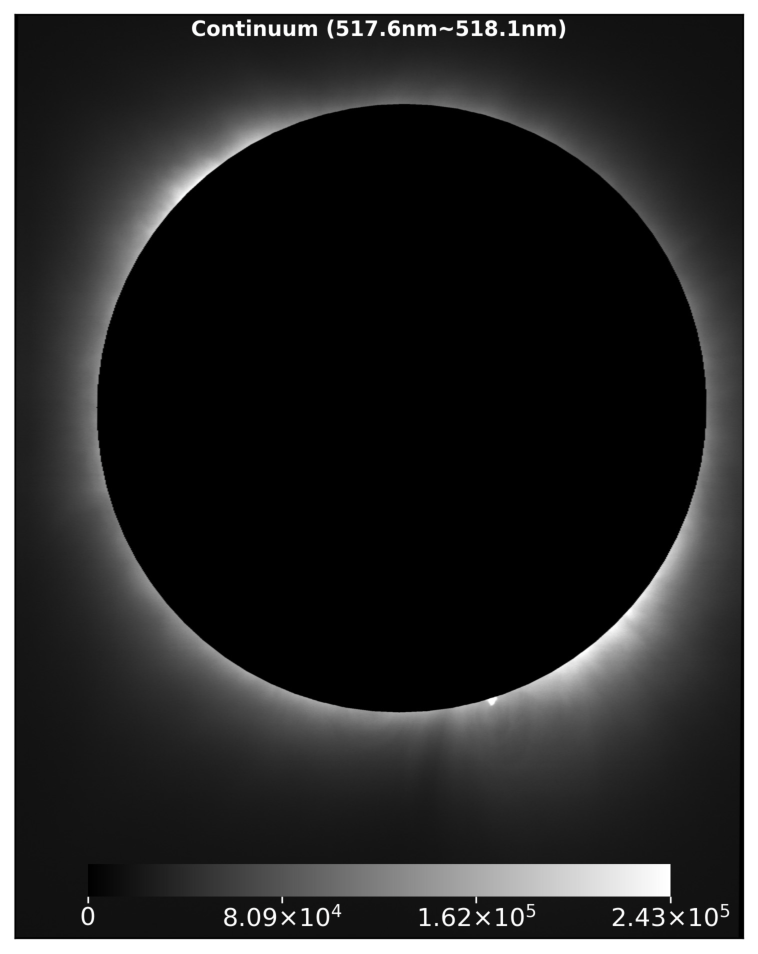}
\hspace{-0.2cm}
\includegraphics[width=5.0cm,height=6.26cm]{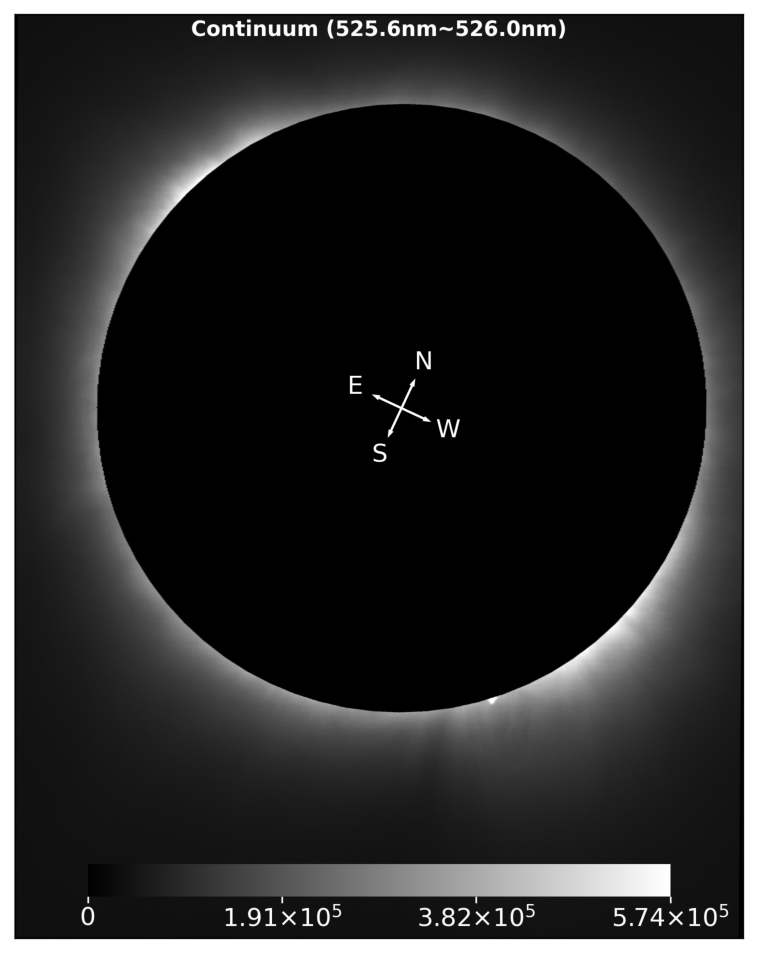}
\hspace{-0.2cm}
\includegraphics[width=5.0cm,height=6.26cm]{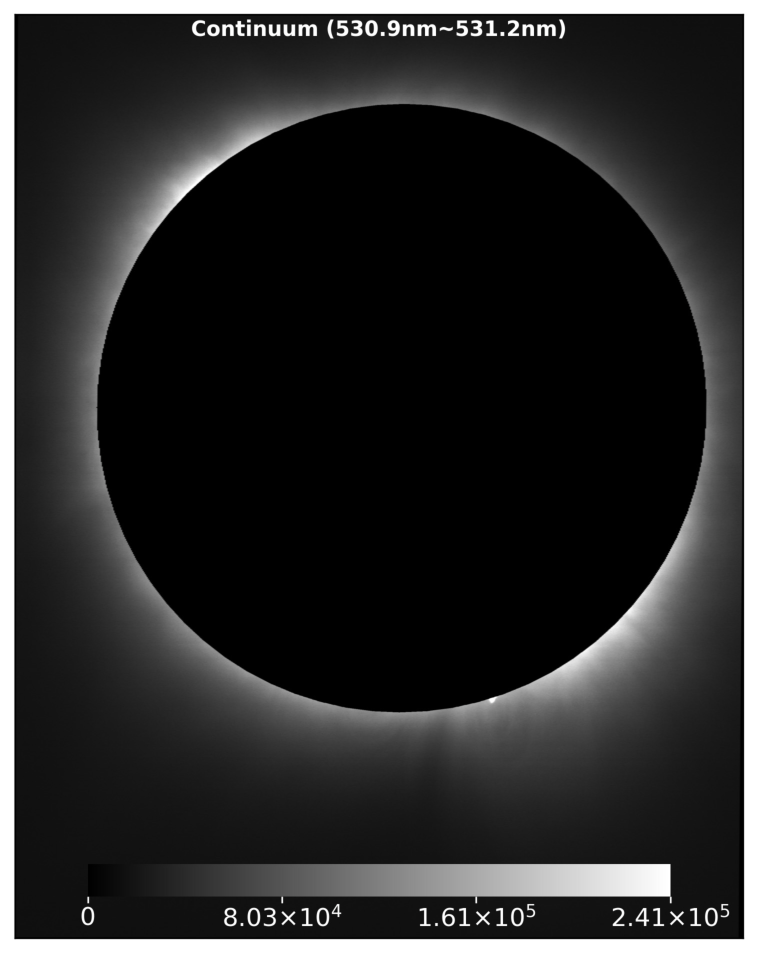}\\
\flushleft\vspace{-0.6cm}\hspace{-1.68cm}
\includegraphics[width=5.0cm,height=6.26cm]{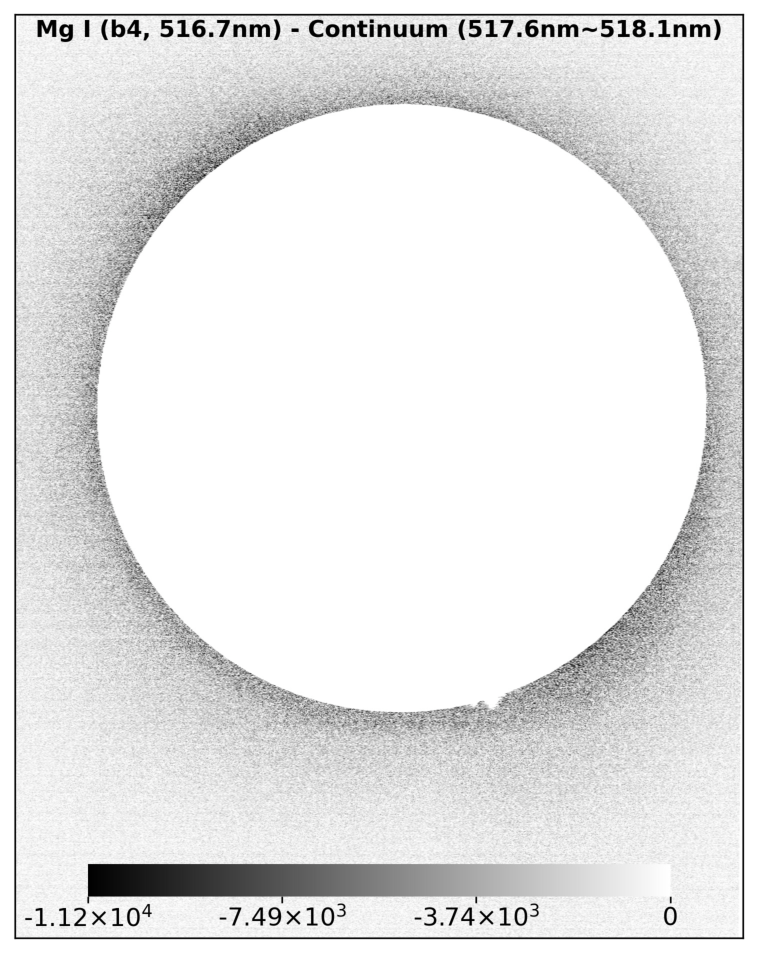}
\hspace{-0.2cm}
\includegraphics[width=5.0cm,height=6.26cm]{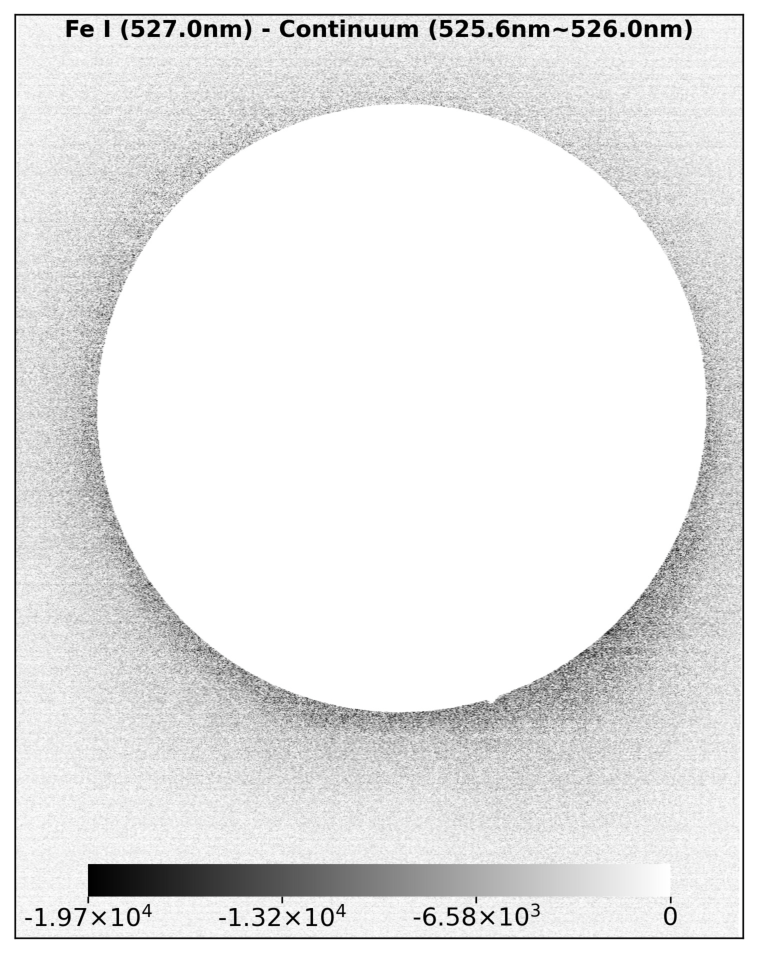}
\hspace{-0.2cm}
\includegraphics[width=5.0cm,height=6.26cm]{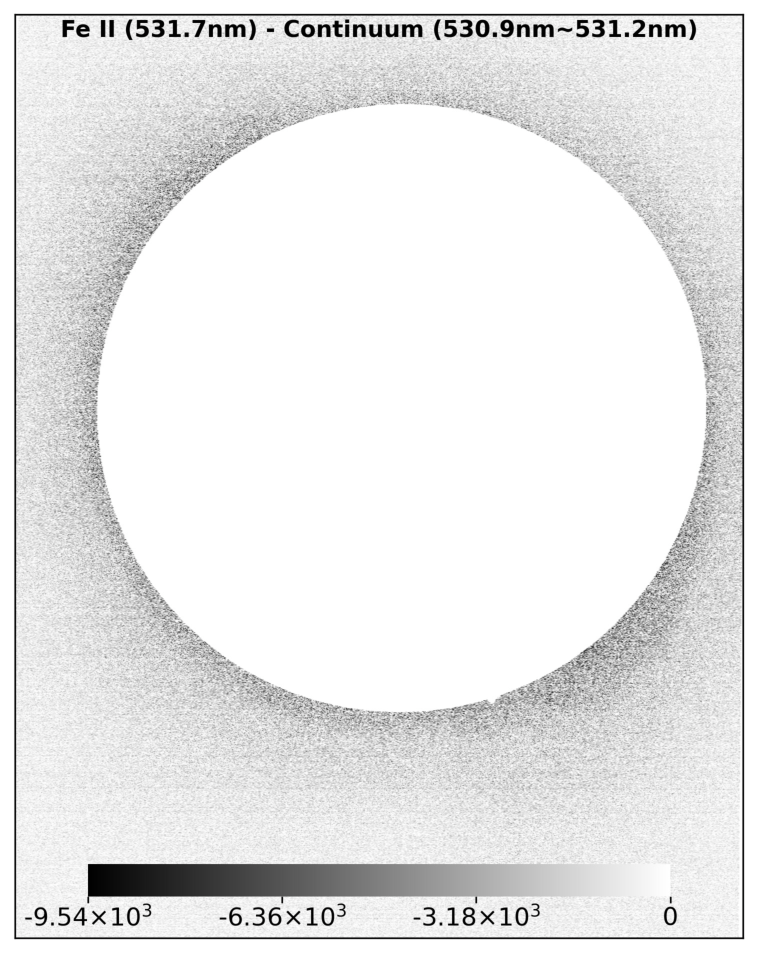}\\
\flushleft\vspace{-0.6cm}\hspace{-1.68cm}
\includegraphics[width=5.0cm,height=6.26cm]{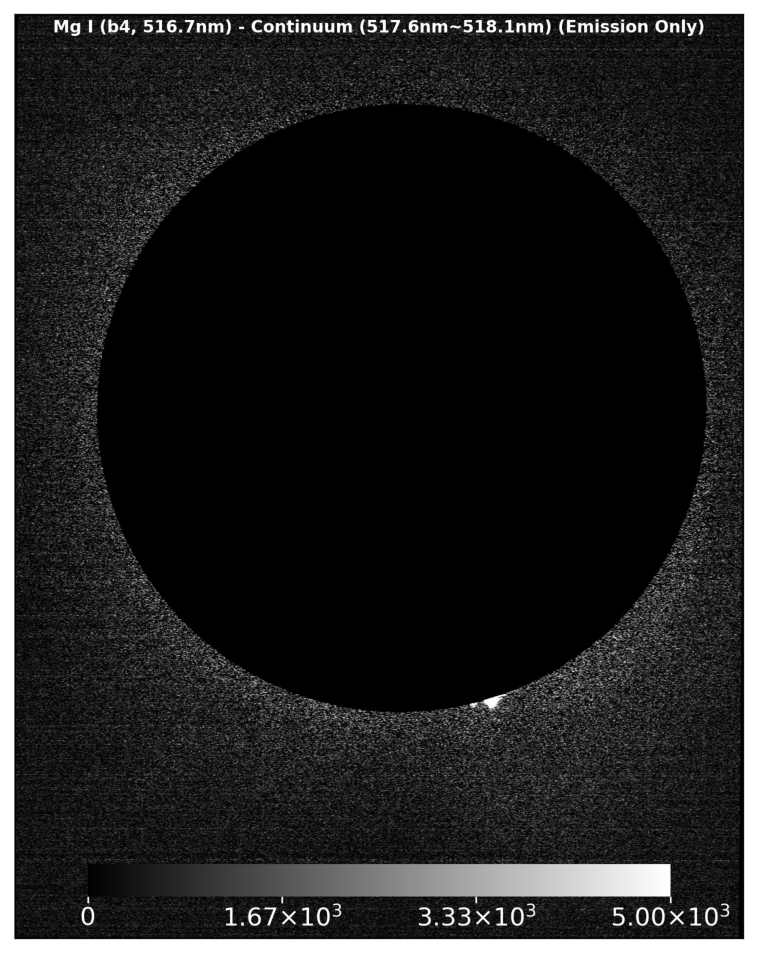}
\hspace{-0.2cm}
\includegraphics[width=5.0cm,height=6.26cm]{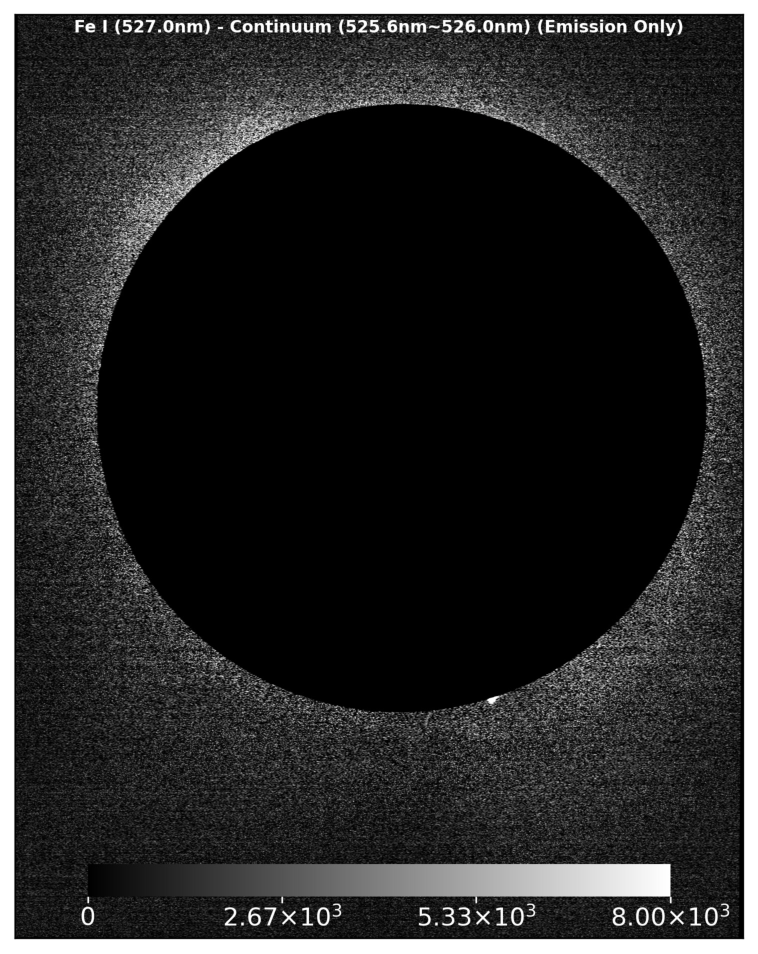}
\hspace{-0.2cm}
\includegraphics[width=5.0cm,height=6.26cm]{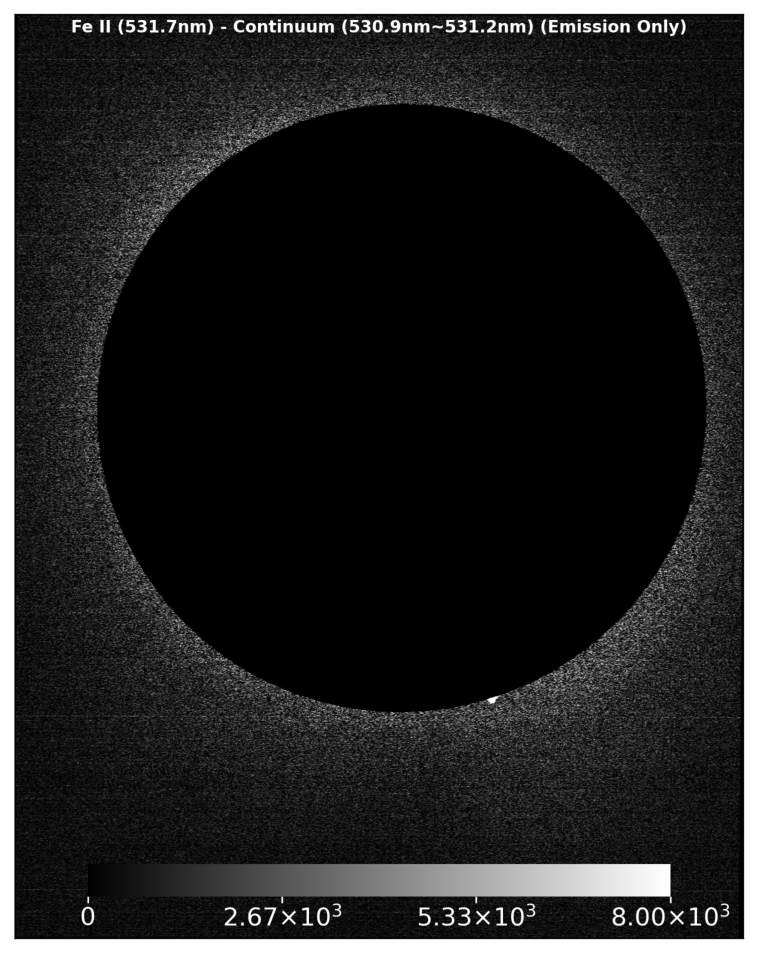}\\
\caption{\footnotesize Maps of original spectral line intensity(top
panels), their adjacent continuum intensity(the second row panels),
the 'monochromatic' F-corona(the third row panels) and intensity of
the counterpart line emission(bottom panels). } \label{}
\end{figure}

Now, let us present the broader band rather than the narrow
'monochromatic' F-corona, plotted in the top left panel of Figure 3,
expressed by the line depressions integrated over the whole
observational band after artificially setting the emission intensity
to be zero again. And map of the line emission caused by the allowed
transitions is also presented after integrations over the eleven
strongest lines including the MgI$b_4$ in the top right panel, via
the line depression artificially setting to be zero. The E-corona is
described here only by the pure green coronal line intensity
obtained after subtracting the corresponding adjacent continuum one,
and further only wavelength points of its line wings are sampled
because those intensities at the line core are saturated in the
detector chip around the solar limb. It seems trial for plotting
K-corona after integration over the whole observational band,
plotted in the bottom right panel, since it looks undistinguishable
from those in the second row maps of Figure 2. But it provides a
reference of intensity magnitude to evaluate magnitudes of the line
depressions in the inner F-corona and the line emission intensities.
It is evident that the signal-to-noise ratio is significantly
improved for the F-corona since depressions of much more lines are
included over the whole band. Again, no internally fine structures
like fibrils can be found either in the F-corona or their line
emission counterpart distribution. This property signifies their
nature and lead to a conclusion that the neutral atom fluxes are not
components of matter in the loops unless they are ionized after
heating. Both the north-south and the east-west asymmetries
remain.Although thicker by one order, the broader band F-corona
looks like the monochromatic F-corona at 531.7nm, while the
integrated line emission distribution looks like that at the MgI
$b_4$ line, with intensity amplitudes much less than depths of the
broad band inner F-corona on the whole. It is evident that the
distribution of these emitters looks more discrete than that of the
inner F-corona, and most of them are unreal but produced by noise.

In fact, both the F-corona and the line emission distribution are
contributed primarily from neutral atoms of iron, magnesium,
titanium, etc., and then secondarily from once ionized ions. Their
formation temperatures range from several thousand to a few ten
thousand Kelvin. They form upward flow of low temperature particles.
As tracers of these neutral atoms and ions, it is easily understood
that these Fraunhofer line depths and line emission intensities are
proportional to number of these scatterers or emitters in the
tenuous coronal atmosphere.

It seems from these three maps that the strong Fraunhofer line
depression and their corresponding emission are more concentrated in
the loops, especially of the active regions. It may mean that the
diffusion is temporarily hindered in these regions because they
become much thinner beyond them. On the contrary, the line
depression and emission concentrations become very thin in the
regions around the northern pole where almost no loops can be seen.
However, such a very small fraction of these diffusion particles
escaped from those loop-dominated regions implies the following
physical scenarios. One is that there should be interaction between
the diffusion of these neutral particles and the loops, and thus
most of the neutral atoms are heated and ionized by thermalization
due to collision of the neutral atoms with the ions and free
electrons in the loops. The other indicates that the newly formed
ions and free electrons after the ionization becomes injected
particles and enhance the density of the charged particles in the
loops. Furthermore, the most of once ionized ions in the fluxes are
also captured by magnetic fields in the loops via Lorentz force. On
the other hand, it indicates that there are passages for a very
small fraction of the neutral atoms and once ionized ions to escape
from the loops, as in the case underneath the corona, so that it is
available for us to see these escaped particles via scattering or
emission detected here.

\begin{figure}
\vspace{1.0cm} \flushleft
\includegraphics[width=4.5cm,height=6.3cm]{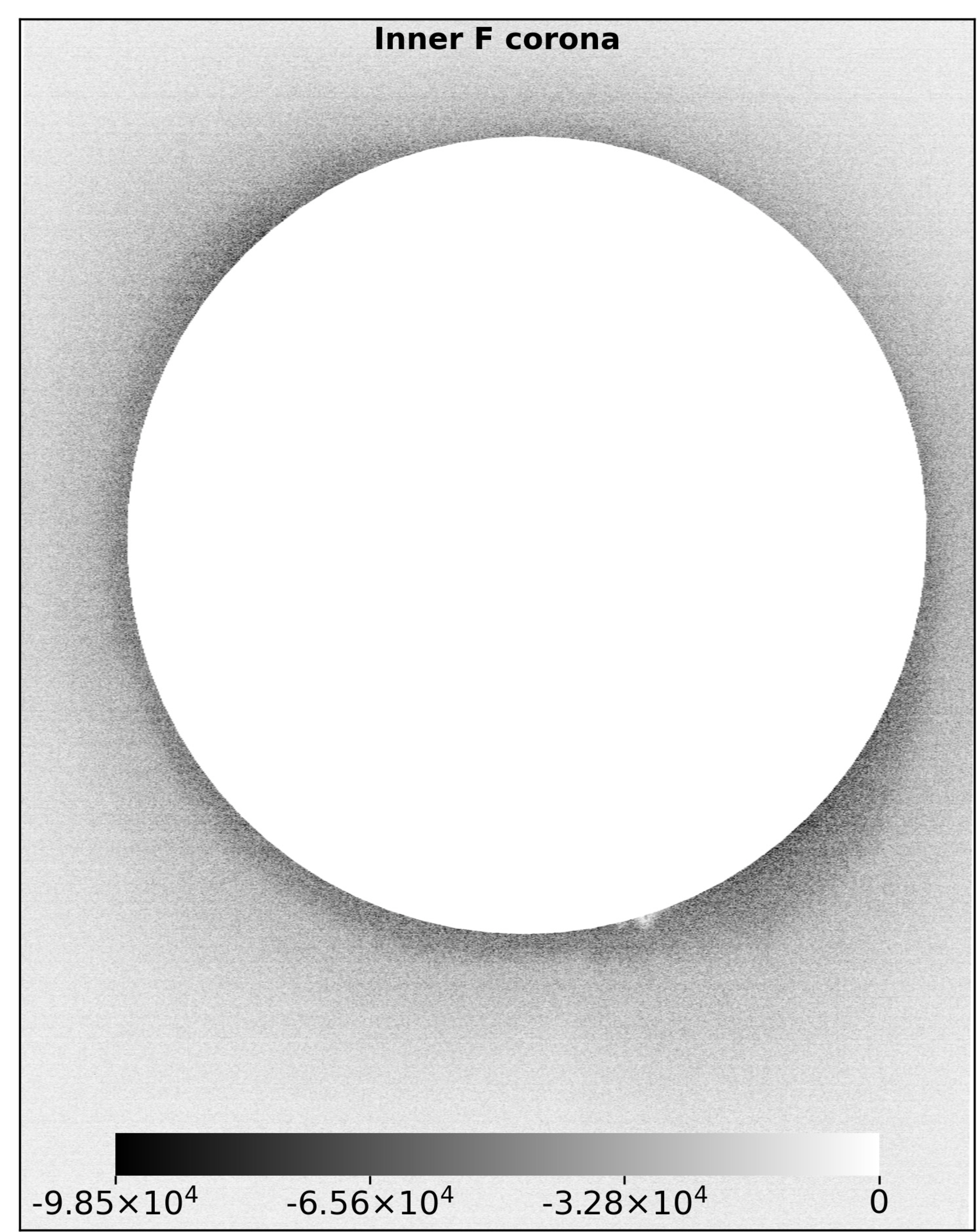}
\hspace{2.0cm}
\includegraphics[width=4.5cm,height=6.3cm]{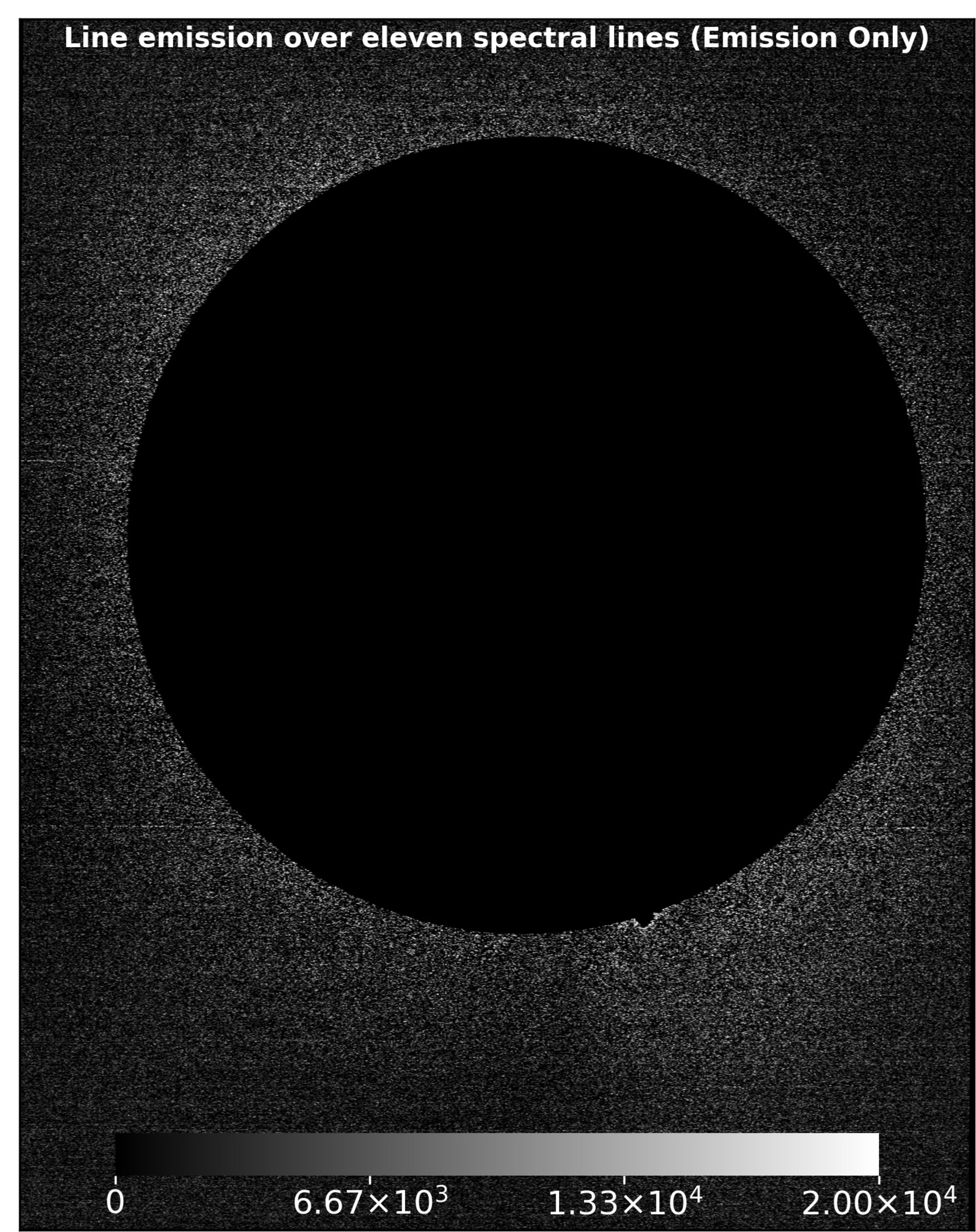}\\
\vspace{1.8cm} \flushleft
\includegraphics[width=4.5cm,height=6.3cm]{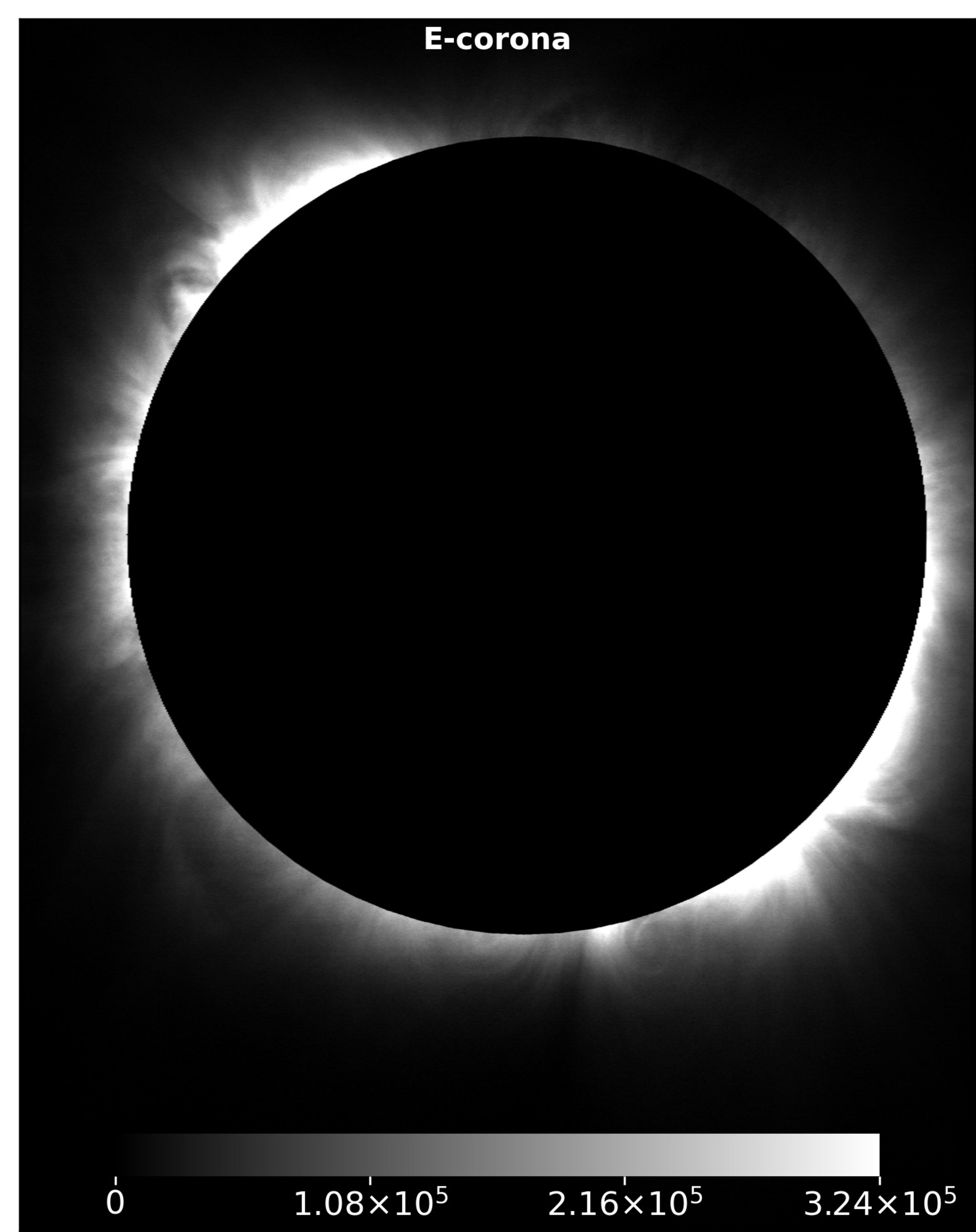}
\hspace{2.0cm}
\includegraphics[width=4.5cm,height=6.3cm]{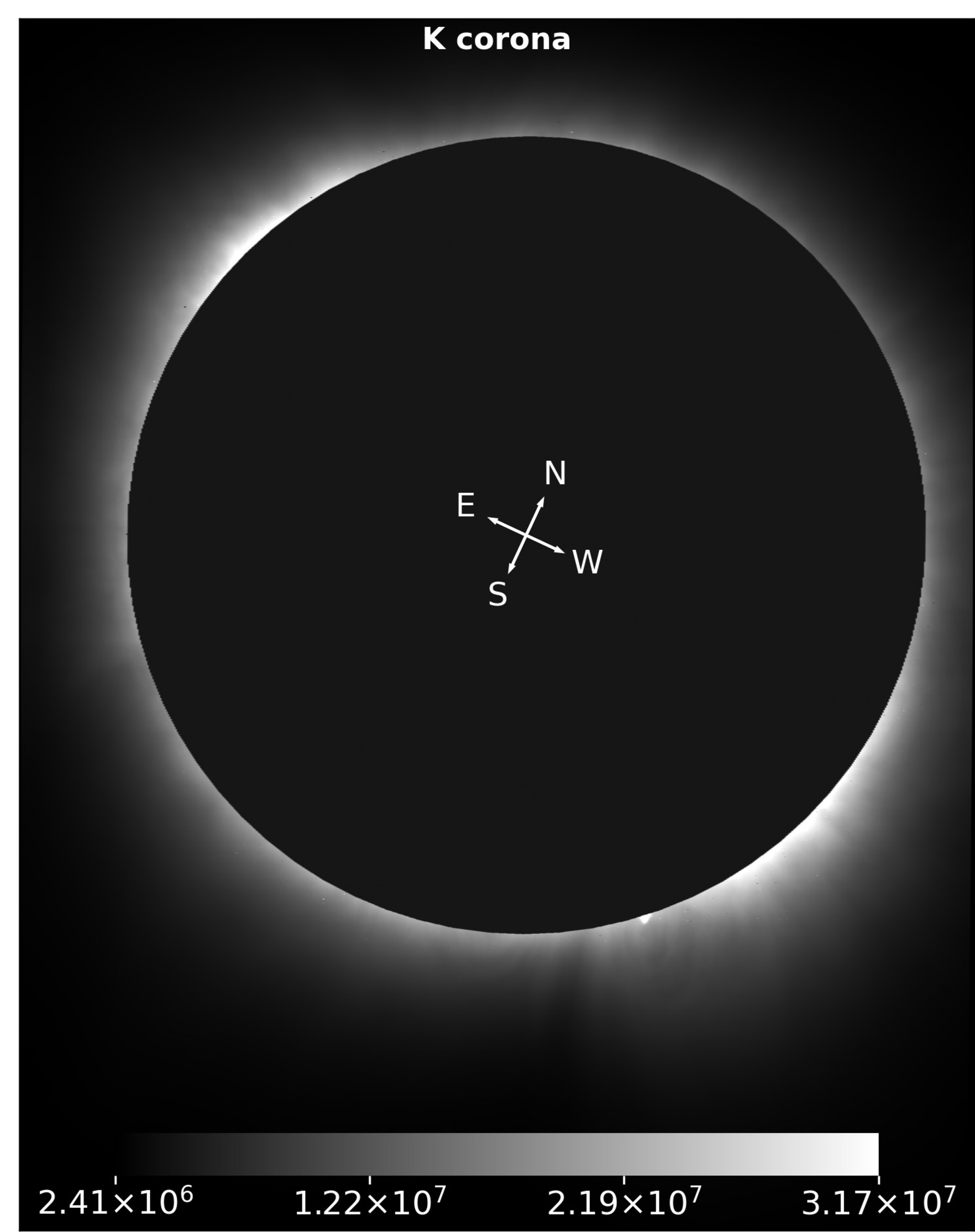}
\caption{\footnotesize Multiple faces of solar corona. Top left: the
inner F-corona representing the Fraunhofer line depression
integrated over the whole observational band. It is formed via
scattering by primarily neutral atoms and fewer once ionized ions.
The diffusion can be witnessed; Top right: map of line emission
intensities contributed from eleven most significant spectral lines
yielded by allowed transitions. It shows more discrete diffusion.
Bottom left: the E-corona generated from the line wings of the
forbidden green coronal line FeXIV530.3nm. It tells us how these
coronal magnetic loops spread and thus contribute to the Cowling
dissipation via collisions of the neutral atom fluxes(indicated as
inner F-corona) with ions in the loops. Bottom right: the K-corona
yielded primarily via scattering by free electrons in the solar
corona, resulted from an integration of all the continuum
intensities within the band for each spatial point in the field of
view. It is much brighter than the other kinds of coronae shown here
by two to three orders as indicated by the grey scale bars plotted
in the bottom part of each panel.} \label{}
\end{figure}

\section{Neutral corona and coronal heating}

The above scenario fits just a picture of Cowling
conductivity\cite{Schluter1950} \cite{Cowling1957}, which is
resulted from the impact of the neutral particles with the ions via
electrodynamical coupling in presence of magnetic field. As a
consequence, the energies of large scales of non-thermal diffusive
motion of the neutral atoms and the current represented by macro
helical motions of constituent ions convert to the thermal one. It
occurs in a way that the neutral flux destroys the ideal
magnetohydrodynamic state into dissipation. From the generalized
Ohm's law, it is proven that the plasma heating rate can become
greater by even eight orders than that resulted from the classical
electron(Spitzer) conductivity in the corona\cite{Zaitsev2008}. In
the present case, the collisions caused by the diffusive neutral
atom flux with the ordered ion motions of the current confined in
the loops lead to the irreversible heating\cite{Klimchuk1992}
\cite{Rosner1978}. In the cases described previously, the percent
line depression defined as ratio of the F-coronal line depth to the
K-coronal intensity peaks at $0.25\%$ and has a median value of
0.30$\%$. According to the analysis by Zaitrev and Stepanov(2008),
just a tiny fraction of the neutral atoms, e.g., $10^{-5}$, injected
into the coronal magnetic loops can lift the resistance by six or
more orders. In fact, the same principle of heating is utilized in
the artificial Tokamak device, known as neutral beam injection(NBI)
to lift the temperature. It has been interpreted in a way like that
described by Ionson\cite{Ionson1984}. Ionson treated the coronal
magnetic loops as parts of RLC circuit, followed by Zaitsev and
Stepanov\cite{Zaitsev2008}, and tried to present a unified theory
emphasizing the electrodynamic coupling efficiency via a general RLC
formalism.

According to the above discussion, global coronal heating efficiency
for the Cowling dissipation depends also on density of the global
distributions of the coronal magnetic loop. We have just shown the
abundance of the loop distribution especially in the E-corona map in
Fig.3. In fact, specific global distribution density depends on the
specific spectral lines with different formation temperatures. In
order to reinforce the wide and dense distributions of the coronal
loops, we choose four SDO/AIA coronal emission maps\cite{Lemen2012}
in Figure 4. They are selected with their acquirement time when the
total eclipse took place. In our opinion, the bright loops at
different spectral lines shown in these SDO/AIA images
\cite{Litwin1993} of Figure 4 illustrate just heating phases in such
an optically thin coronal atmosphere. It should be noted that the
directions for these maps are different from those in our eclipse
maps shown in Figures 2 and 3, as shown in the bottom left corner of
top right panel of Figure 4.

Now, let us detail the information contained in these maps. The top
left map in Figure 4 is derived at the once ionized helium
HeII30.4nm line with formation temperature of 5.0$\times
10^{4}K$\cite{Lemen2012}. It is not difficult to find a limited
number of coronal magnetic loop systems as sets of the brightest
arches over an active region above the northeastern limb as well as
on the disk. It is easy to see that much more loops appear in map at
17.1nm(top right panel). But the most abundant loops can be
witnessed in map at 21.1nm line(bottom left panel). However, map at
94.0nm line(bottom right panel) shows weaker loops though its
formation temperature is the highest. According to Lemen and
cooperators\cite{Lemen2012}, for FeIX 17.1nm line, the temperature
in favor of its formation is at $6.3\times 10^{5}K$. Temperature
$2.0\times 10^{6}K$ is the most suitable for yielding FeXIV 21.1nm
line. Finally, the formation temperature of FeXVIII9.4nm line stands
at 6.31$\times 10^{6}K$. The 21.1nm map shows not only the widest
and densest coronal loops on the solar disk as well as above the
limb, but also their surrounding mosses. The mosses may provide a
clue about the heat dispersal via cross-field motion during the loop
magnetic flux decays, due to the abnormal resistivity or
overpressure in the heated loops and/or magnetic field inhomogeneity
of the loops\cite{Litwin1993}. The increase of densities can be
expected due to generation of new free electrons and ions after
ionization of the neutral atoms by the heating, resulted in the
plasma injection into these neighboring loops. Therefore, coronal
loop heating could be the basis for the coronal heating.

These different coronal appearances in Figures 3 and 4 reflect the
different thermal states thus probably different heating phases,
because the thermal dynamical equilibrium cannot be effectively
established due to too thin atmosphere. It is noteworthy that in
these maps not all the coronal loops are included because gaps
exists among their formation temperatures, and the lifetimes of the
coronal loops of the quiet sun and the coronal holes are more than
tens of minutes\cite{Aschwanden2015} \cite{Kraus2025}, much longer
than the total eclipse lasting duration.

Now, let us go further. As mentioned above, a tiny fraction of the
neutral atoms and once ionized ions escape still from heating in the
coronal loops as the thinnest line depression part above the
thickest. This scenario reminds us that similar physical processes
may also happen in the chromosphere and transition zone, most of
which are filled with smaller magnetic loops\cite{Madjarska2023}
\cite{Nobrega-Siverio2023} \cite{Judge2024a}. Though the Cowling
dissipation can be more easily traced in the corona due to the
'frozen effect' ascribed to much lower ratio $\beta$ of the plasma
pressure to the magnetic one, i.e., $\beta\ll 1$, it could also come
into play in the atmospheric layers beneath the corona with denser
neutral atoms and magnetic loops with stronger field strengths
resulting in $\beta$$<= 1$ \cite{Wang1993} \cite{Yalim2020}. As
pointed out by Aschwanden\cite{Aschwanden2015}, the heating rate is
much more dependent on the particle density than the loop length.
The efficiencies are also based on observational fact that
emergences of 95$\%$ photospheric magnetic fluxes form the 'magnetic
carpet' or canopy tops below the corona, and only about 5$\%$ fluxes
give births to large-scale coronal loops\cite{Priest2002}
\cite{Rempel2014} \cite{Lites2017}. This makes the Cowling
dissipation more effective in the chromosphere\cite{Wang1993}
\cite{Yalim2020} and even most efficient in the transition zone,
thus leads to the sudden rise of temperature.

\begin{figure}
\vspace{3cm}\flushleft\hspace{-0.4cm}
\includegraphics[width=1.0cm,height=1.46cm]{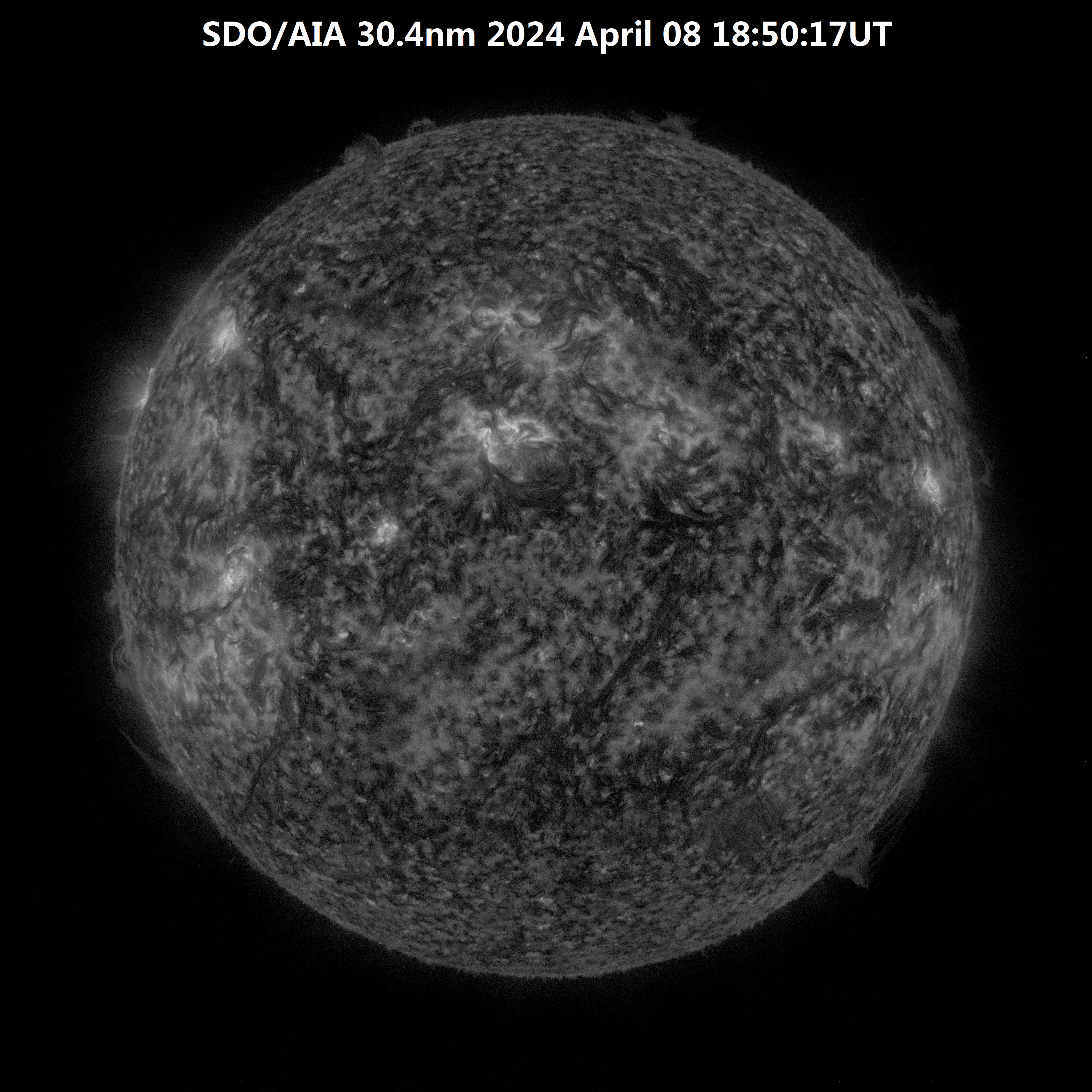}
\hspace{6.2cm}
\includegraphics[width=1.68cm,height=2.43cm]{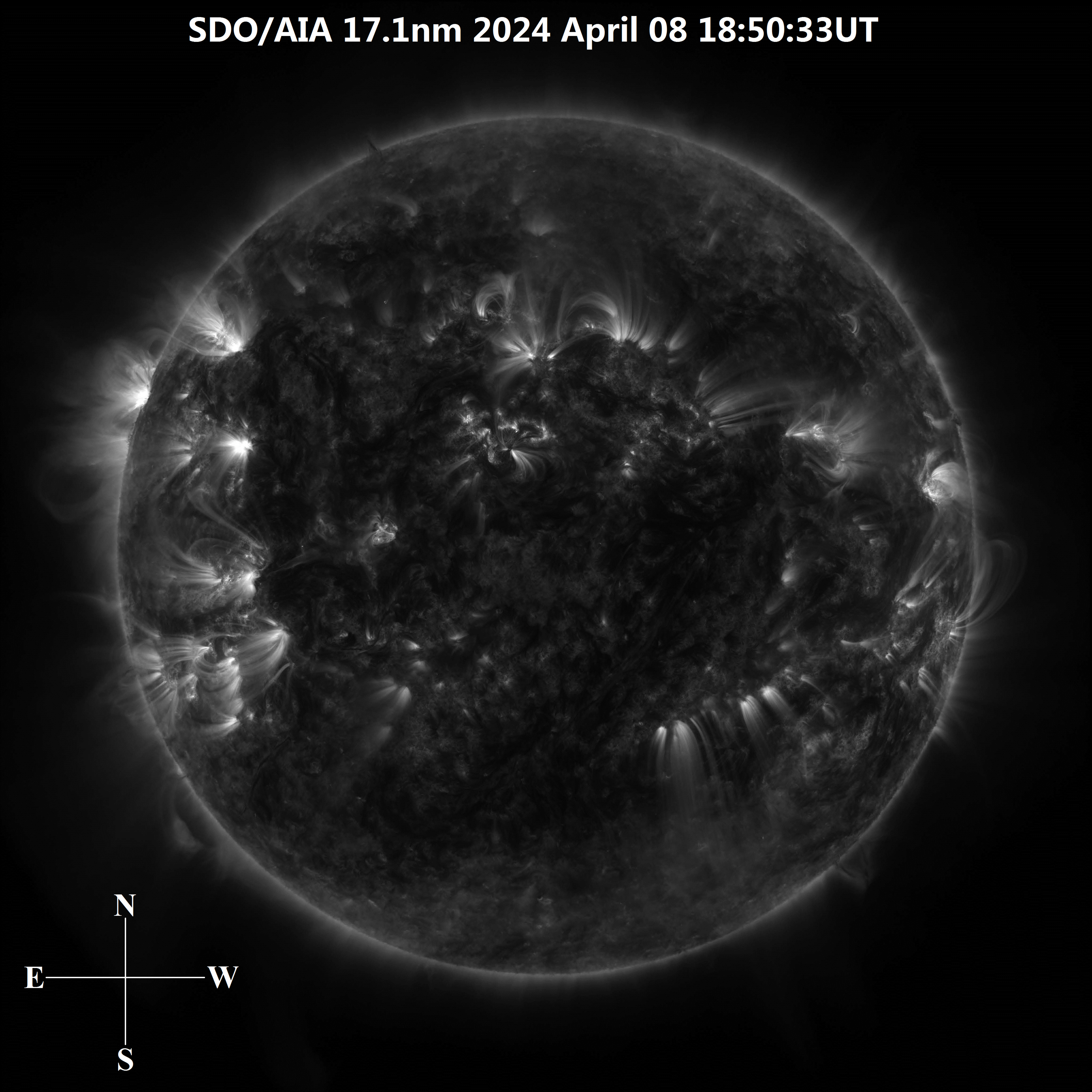}\\
\vspace{4.4cm} \flushleft\hspace{-0.4cm}
\includegraphics[width=1.0cm,height=1.46cm]{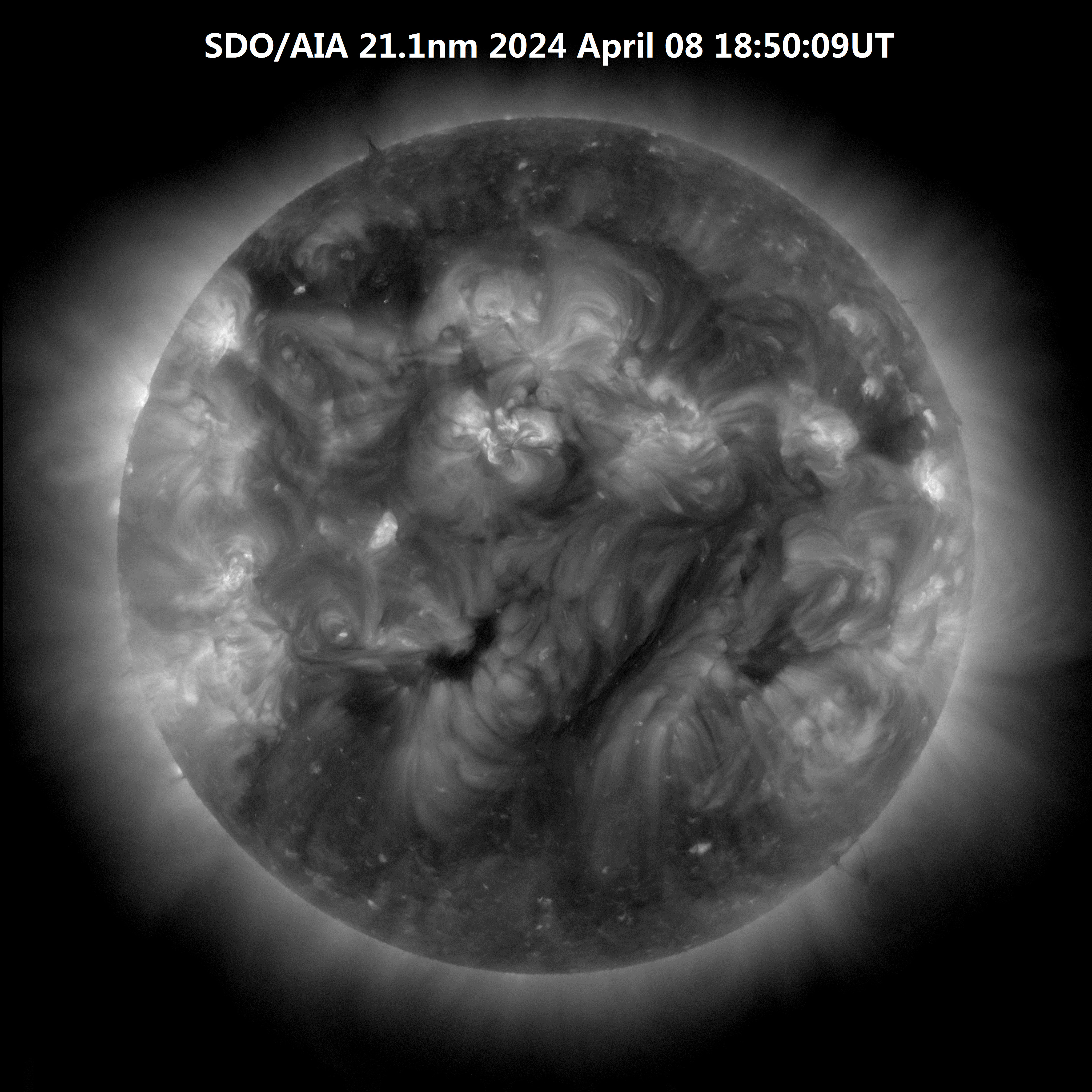}
\hspace{6.2cm}
\includegraphics[width=2.01cm,height=2.92cm]{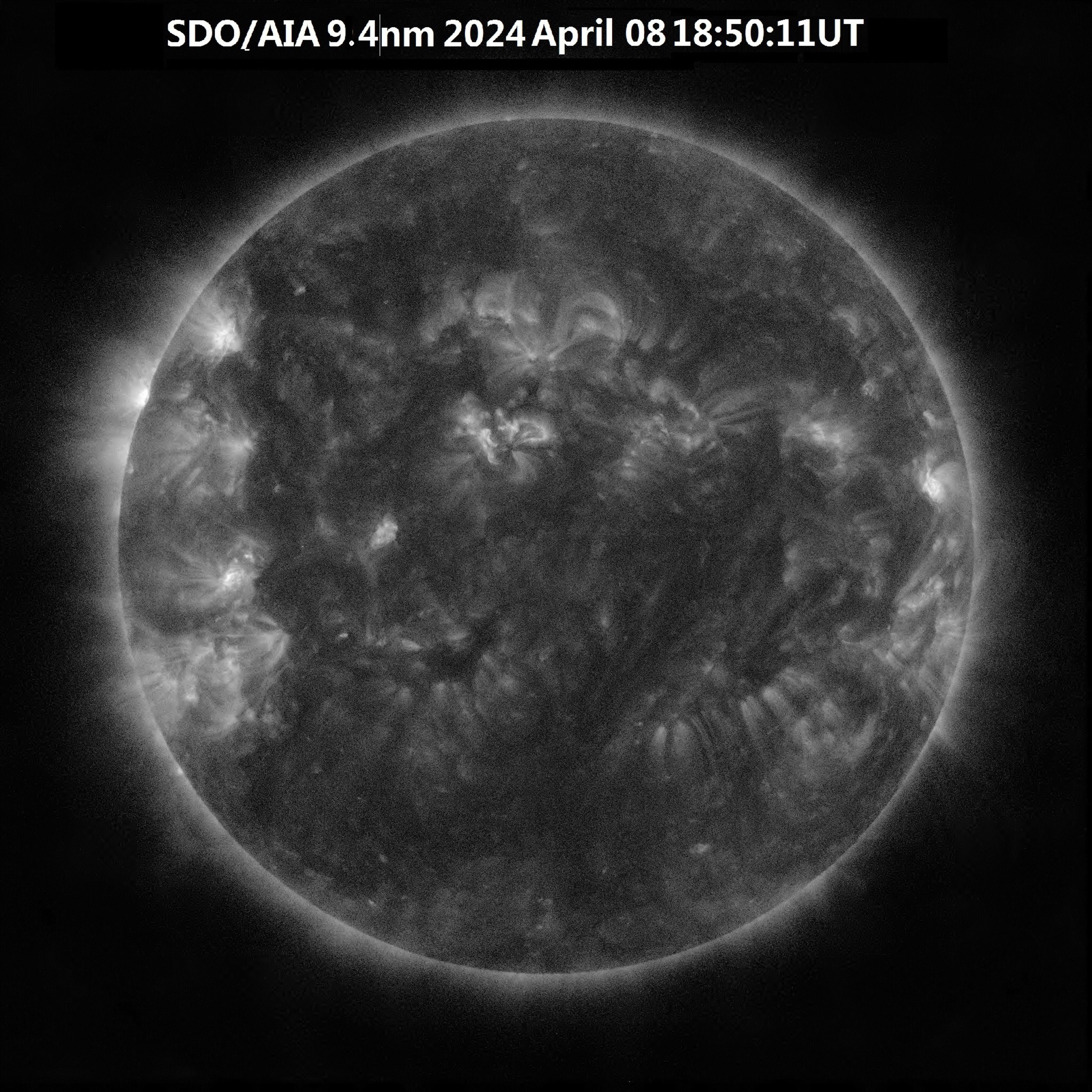}
\caption{\footnotesize Solar corona with multiple faces reflecting
its multi-thermal essence and probably hinting different coronal
heating phases by the abnormal Cowling resistance. These coronal EUV
maps were obtained from SDO/AIA respectively at 30.4nm(top left),
17.1nm(top right), 21.1nm(bottom left) and 9.4nm(bottom right). All
the maps were selected of acquirement time when the total solar
eclipse occurred. Solar corona in these maps corresponding to
different formation temperatures increasing from the left to right
and from the top to bottom. As the temperature increases, more and
more coronal loops were detected, and mosses become more clear till
the formation temperature of FeFeXIV21.1nm. And then they become
weaker at the formation temperature of FeXVIII9.4nm(bottom right). }
\label{}
\end{figure}

\section{Discussions and Conclusions}

According to the above discussion, concatenation containing
magneto-convection in the photosphere, the emergence of magnetic
flux forming the magnetic loops and the diffusion of the neutral
atoms, along with energy losses from radiation, conduction and solar
wind in the corona, provides us a clear clue to form a continuous
and unified scenario about the primary heating processes along
atmospheric height from the chromosphere and transition zone to the
corona via heating cascades. The particles outflow from the
photosphere can be divided into two parts for heating. The most part
is heated via Cowling dissipation due to the collisions of ions with
the neutral atoms in small magnetic loops in the layers below the
corona. In the magnetic carpet with complex magnetic field
configuration, it is believed that a great amount of current sheets
continue to form via flux tube tectonics from magneto-convection in
the photosphere and then dissipate\cite{Priest2002}
\cite{Priest2011}. Therefore, it seems reasonable to conjecture that
the Cowling effect could take effect with the upward neutral atom
flux in the magnetic carpet, where the temperature could be raised
very abruptly to form the transition region with thickness close to
that of the carpet, due to much stronger magnetic field strength yet
tenuous plasma in case of $\beta \le 1$. In fact, the presence of
the neutral flux can speed up the diffusion of magnetic loops by
eight to ten orders\cite{Zaitsev2008}. The other part is just the
neutral atoms and once ionized ions escaped from the heating in the
magnetic carpet, reported here. Related to power supply from the
magnetic convection of the loop footpoints in the photosphere and
cooling by radiative loss, heat conduction and even solar winds, the
coronal heating concatenation containing the neutral atom diffusion
along with the intermittent emergence of magnetic
flux\cite{Ryutova2006} \cite{Leake2006} \cite{Magara2012} forms a
chain of temperature self-regulation in the solar atmosphere.

Besides the coronal loops which show simple magnetic field
configurations, another kind of sites provides also a platform for
occurrence of the Cowling dissipation. That is the current sheet
around special topological magnetic configurations called
separatrices(e.g., Priest et al.\cite{Priest2002}). As the primary
component of the inner F-corona, the outward flux of the neutral
atoms can permeate into the current sheets to promote critically
Joule dissipation in the same way described previously, thus enhance
greatly the heating efficiency via both the slow and fast
reconnections\cite{Uzdensky2007} \cite{Ni2012} \cite{Yalim2020}
\cite{Wargnier2023}. It is valuable to note that microflares and
nanoflares due to the small-scale magnetic reconnections were
proposed by Parker\cite{Parker1972} \cite{Parker1988} to be a main
mechanism for the coronal heating, and then, for instance, developed
by Priest and his cooperators via a more detailed coronal tectonics
model\cite{Priest2002} \cite{Priest2011}. The content of the model
includes build-up of current sheets attributed to motions of
photospheric footpoints along the separatrix boundaries of the flux.
Therefore, the neutral atom flux or neutral corona plays roles again
in both the resistive and reactive ways of the coronal heating in
the thermodynamical model, described by the routines of Fig.2.4 of
Judge and Ionson\cite{Judge2024b}, where the coronal magnetic loops
and their induced coronal electric currents play the core role with
Poynting flux emerged from the photosphere due to the
magneto-convection and magnetic flux emergence\cite{Leake2006}
\cite{Huang2018} \cite{Judge2024b}.

Finally, the efficiency of heating may be further improved by taking
into account of dissipation of energy of the
magnetohydrodynamic(MHD) waves such as Alfv$\acute{e}$n and kink
waves, triggered by the neutral flow. The loops are really sensitive
wave resonators\cite{Zaitsev2008} \cite{Priest2011}. The dissipation
of the MHD waves is also regarded as a main mechanism responsible
for the coronal heating(e.g., van Ballegooijen, et al.
\cite{Ballegooijen2017}). Therefore, the global distributions of the
neutral atom fluxes and coronal loops of different scales can unite
the theory of the coronal heating as the results of cascade of
heating events from the chromosphere to the corona.

\vspace{0.6cm}

{\it This work is sponsored by National Science Foundation of China
(NSFC) under the grant numbers 11527804 and U1931206. The authors
thank the Solar Dynamic Observatory(SDO) consortium for SDO/AIA
pictures used in this paper.}


\begin{thebibliography}{99}
\bibitem{Grotrian1934} Grotrian, Von W., '$\ddot{U}$ber das Fraunhofersche Spektrum der Sonnenkorona', {\it
Zeitschrift f$\ddot{u}$r Astrophysik}, {\bf 8},124-137(1934)
\bibitem{Edlen1943} Edl$\acute{e}$n, Bengt, 'Die Deutung der Emissionslinien im Spektrum der Sonnenkorona.
Mit 6 Abbildungen', {\it Zeitschrift fur Astrophysik}, 22,
30-64(1943)
\bibitem{Aschwanden2015} Aschwanden, M.J., 'Physics of the solar
corona', {\it Springer, Published in association with Praxis
Publishing Chichester, UK},  2015, {ISBN: 3-540-22321-5}
\bibitem{Judge1998} Judge, D. L., McMullin, D. R., Ogawa, H. S., et
al., 'First solar EUV irradiances obtained from SOHO by the
CELIAS/SEM', {\it Solar Physics}, 177, 161-173(1998)
\bibitem{Stellmacher1974} Stellmacher, G. and Koutchmy, S., 'Study
of low dispersion eclipse spectra: Observation of weak low
extinction emission lines in the corona', {\it Astronomy $\&$
Astrophysics}, 35, 43-48(1974)
\bibitem{Judge2004} Judge, P.G., and Pietarila, A.,  'On the formation of
extreme-ultraviolet helium lines in the sun: analysis of SOHO data',
{\it the Astrophysical Journal}, 606:1258-1275(2004)
\bibitem{Moise2010} Moise, E.,  Raymond, J., and Kuhn, J. R.,
'Properties of the diffuse neutral helium in the inner heliosphere',
{\it the Astrophysical Journal},722:1411-1415(2010)
\bibitem{Morgan2007} Morgan, H., and Habbal, S.R., 'The long-term stability of the visible F corona at heights
of 3-6 R$_{\bigodot}$', {\it Astronomy $\&$ Astrophysics},
471:L47-50(2007)
\bibitem{Stenborg2021} Stenborg, G., Howard, R.A., Hess, P.,
Gallagher, B., 'PSP/WISPR observations of dust density depletion
near the Sun. I. Remote observations to 8 Rsun from an observer
between 0.13 and 0.35 AU', {\it Astronomy $\&$ Astrophysics},
650:A28(10pp)(2021)
\bibitem{Boe2021} Boe, B., Habbal, S., Downs, C., and Druckmueller, M.,
'The color and brightness of the F-corona inferred from the 2019
July 2 total solar eclipse', {\it the Astrophysical Journal},
912:44(15pp)(2021)
\bibitem{Burtovoi2022} Burtovoi, A. Naletto, G., Dolei, S., Spadaro, D., Romoli, M., F. Landini, F.,
and De Leo, Y., 'Measuring the F-corona intensity through time
correlation of total and polarized visible light images', {\it
Astronomy $\&$ Astrophysics}, {\bf 659}, A50(16 pages)(2022)
\bibitem{Lamy2022} Lamy, P.L., Gilardy, H., and LIebaria, A.,
'Observations of the solar F-corona from space', {\it Space Sci.
Rev.}, {\bf 218}, 53(72 pages)(2022)
\bibitem{Russell1929} Russell, H.N., 'On meteoric matter near the stars', {\it the Astrophysical Journal},
69,49-71(1929)
\bibitem{Qu2024} Qu, Z.Q., Chang, L., Dun, G.T., X.M. Cheng, et al.,
'Spectropolarimetry of Fraunhofer lines in local upper solar
atmosphere', {\it the Astrophysical Journal}, 974:63(13pp)(2024)
\bibitem{Qu2011} Qu, Z.Q., 'A fiber arrayed solar optical
telescope', Solar Polarization 6, ASP Conference Series,
Vol.437,423-431(2011), Kuhn, Berdyugina, Harrington, Keil, Lin,
Rimmele, and Trujillo-Bueno, eds.
\bibitem{Qu2014} Qu, Z.Q., Chang, L., Cheng, X. M., et al., 'Prototype
FASOT', Solar Polarization 7, ASP Conference Series, Vol.489,
263-270(2014), K. N. Nagendra, J. O. Stenflo, Zhongquan Qu, and M.
Sampoorna, eds.
\bibitem{Deutsch1964} Deutsch, A.G. and Righini, G., 'An airborne observation of the
coronal spectrum at the eclipse of July 20, 1963', {\it the
Astrophysical Journal}, 140, 313-318(1964)
\bibitem{Kumar2023} Kumar, H., Kumar, B., Rajaguru, S.P., et al., 'A study of the propagation
of magnetoacoustic waves in small-scale magnetic fields using solar
photospheric and chromospheric Dopplergrams: HMI/SDO andMAST
observations', {\it Journal of Atmospheric and Solar-Terrestrial
Physics}, 247-254(2023)
\bibitem{Yu2024} Yu, F., Rao, S., Zhao, J., et al.,
'Source region of the solar wind: Statistics of the Doppler
velocities at the chromosphere', {\it the Astrophysical Journal
Letters}, 968: L20(11pp),2024
\bibitem{Tian2021} Tian, H., Hara, L., Baker, D., et al., 'Upflows in the upper solar
atmosphere', {\it Solar Physics}, 296, 47(2021)
\bibitem{Schluter1950} Schluter, A., and Biermann, L.Z.,
'Interstellare magnetfelder', {\it Naturforsch}, A5, 237(1950)
\bibitem{Cowling1957} Cowling, T.G., {\it Magnetohydrodynamics}, Interscience Publishers, New York, 1957
\bibitem{Zaitsev2008} Zaitsev, V, and Stepanov, A.V., 'Coronal magnetic
loops', {\it Physics-Uspekhi}, 51, 1123-1160(2008)
\bibitem{Klimchuk1992} Klimchuk, J.A., Lemen, J.R., Feldman, U.,et al.,
'Thickness Variations along Coronal Loops Observed by the Soft X-Ray
Telescope on YOHKOH', {\it Publications of the Astronomical Society
of Japan}, 44, L181-L185(1992)
\bibitem{Rosner1978} Rosner, R., Golub, L., Coppi, B., and Vaiana, G.
S, 'Heating of coronal plasma by anomalous current dissipation',
{\it the Astrophysical Journal}, 222, 317-332(1978)
\bibitem{Ionson1984} Ionson, J.A., 'A unified theory of
electrodynamic coupling in coronal magnetic loops: the coronal
heating problem', {\it Astrophysical Journal}, 276:357-368(1984)
\bibitem{Lemen2012} Lemen, J.R., Title, A.M.,
Akin, David J., et al., 'The Atmospheric Imaging Assembly (AIA) on
the Solar Dynamics Observatory (SDO)', {\it Solar Physics}
275:17-40(2012)
\bibitem{Litwin1993} Litwin, C., and Rosner, R., 'On the structure of
solar and stellar coronae: loops and loop heat transport', {\it the
Astrophysical Journal}, 412:375-385(1993)
\bibitem{Kraus2025} Kraus, I., Bourdin, Ph.-A., Zender, J., et al.,
'Coronal bright point statistics: II. Magnetic polarities and mini
loops', {\it Astronomy $\&$ Astrophysics}, 694, A240(7pp)
\bibitem{Madjarska2023} Madjarska, M., Galsgaard, K., and Wiegelmann,
T., 'Photospheric magnetic flux and coronal emission properties of
small-scale bright and faint loops in the quiet sun', 2023, {\it
Astronomy $\&$ Astrophysics}, 678, A32
\bibitem{Nobrega-Siverio2023} Nobrega-Siverio, D., Moreno-Insertis, F., Galsgaard,
K., Krikova, K., and van der Voort, L.P., 'Deciphering solar coronal
heating: Energizing small-scale loops through surface convection',
{\it the Astrophysical Journal}, 958:L38(8pp)(2023)
\bibitem{Judge2024a} Judge, P.G., and Kuin, N.P.M., 'On the intermittency
of hot plasma loops in the solar corona', {\it the Astrophysical
Journal}, 970:130(9pp)(2024)
\bibitem{Wang1993} Wang Jingxiu, 1993, 'Electric conductivity of
lower solar atmosphere', {\it ASP Conference Series}, 46, 465-468
\bibitem{Yalim2020} Yalim, M.S., Prasad, A., Pogorelov, N.V., et
al., 'Effects of Cowling resistivity in the weakly ionized
chromosphere', {\it the Astrophysical Journal}, 899: L4(7pp)(2020)
\bibitem{Priest2002} Priest, E.R., Heyvaerts, J.F., Title, A.M. ,
'A flux-tube tectonics model for solar coronal heating driven by the
magnetic carpet', {\it the Astrophysical Journal}, 576,
533-551(2002)
\bibitem{Rempel2014} Rempel, M., 'Nonlinear simulations of quiet sun
magnetism: on the contribution from a small-scale dynamo', {\it the
Astrophysical Journal}, 789, 132(2014)
\bibitem{Lites2017} Lites, B.W., Rempel, M., Borrero, J.M., et al., 'Are internetwork magnetic
fields in the solar photosphere horizontal or vertical?', {\it the
Astrophysical Journal}, 412, 14(2017)
\bibitem{Priest2011} Priest, E.R., 'The flux tube tectonics
model for coronal heating', {\it Journal of atmospheric and
solar-terrestial physics}, 73, 271-276(2011)
\bibitem{Ryutova2006} Ryutova, M. and Shine, R., 'Coupling effects throughout the
solar atmosphere: Emerging magnetic flux and structure formation',
{it Journal of Geophysical Research}, 111, A03101(10pp)(2006),
\bibitem{Leake2006} Leake, J.E. and Arber, T.D., 'The emergence of magnetic flux through a partially ionised solar
atmosphere', {\it Astronomy $\&$ Astrophysics}, 450, 805-818(2006)
\bibitem{Magara2012} Magara, T., 'How Much does a magnetic flux tube emerge into the Solar
atmosphere?', {\it the Astrophysical Journal}, 748, 53(7pp)(2012)
\bibitem{Uzdensky2007} Uzdensky, D.A., 'Self-regulation of solar
coronal heating process via the collisionless reconnection
condition', {\it Physical Review Letters}, 99, 261101(2007)
\bibitem{Ni2012} Ni Lei, Roussev, I.I.; Lin, Jun; Ziegler, U.,
'Impact of temperature-dependent resistivity and thermal conduction
on plasmoid instabilities in current sheets in the solar corona',
{\it the Astrophysical Journal}, 758:20(11pp)(2015)
\bibitem{Wargnier2023} Wargnier, Q.M., Martinez-Sykora, J.,
Hansteen, V.H., and De Pontieu, B., 'Multifluid simulations of
upper-chromospheric magnetic reconnection with helium-hydrogen
mixture', {\it the Astrophysical Journal}, 946:115(24pp)(2023)
\bibitem{Parker1972} Parker, E.N., 'Topological dissipation and
the small-scale fields in turbulent gases', {\it the Astrophysical
Journal}, 174, 499-510(1972)
\bibitem{Parker1988} Parker, E.N., 'Nanoflares and the solar X-ray corona',
{\it the Astrophysical Journal}, 330, 474-479(1988)
\bibitem{Judge2024b} Judge, P.J. and Ionson, J.A., 'The problem of
coronal heating', {\it Astrophysics and Space Science Library},
Volume 470, Springer, ISBN 978-3-031-46272-6(2024)
\bibitem{Huang2018} Huang, Zhenghua, 'Magnetic Loops above a small flux-emerging region observed by
IRIS, Hinode, and SDO', {\it the Astrophysical Journal},
869:175(13pp)(2018)
\bibitem{Ballegooijen2017} van Ballegooijen, A.A., Asgani-Targhi, M.,
and Voss, A., 'The heating of solar coronal loops by
Alfv$\acute{e}$n wave turbulence', {\it the Astrophysical Journal},
849:46(23pp)(2017)
\end{thebibliography}
\end{document}